\documentclass[twocolumn,english,aps,preprint,prl,reprint,longbibliography]{revtex4-1}
\usepackage[T1]{fontenc}
\usepackage[latin9]{inputenc}
\usepackage{amsmath}
\usepackage{amssymb}
\usepackage{graphicx}
\usepackage{natbib}
\usepackage{babel}
\usepackage{soul}
\makeatletter
\usepackage[usenames,dvipsnames]{color}
 \@ifundefined{textcolor}{}
 {%
   \definecolor{BLUE}{rgb}{0,0,1}
   \definecolor{RED}{rgb}{ 0.6328125,0,0}
 }

\newcommand{\ket}[1]{\left|#1\right>}

\makeatother

\newcommand{\hamiltonian}
{
\begin{equation}\label{Hamiltonian}
	\mathcal{H} = \sum_{i=1}^9 \delta_i \hat{n}_i + 
			      \frac{\eta_i}{2} \hat{n}_i \left( \hat{n}_i - 1 \right) + 
			      \sum_{i=1}^8 g_{i} \left( \hat{a}_{i}^{\dag} \hat{a}_{i+1} + \hat{a}_{i} \hat{a}_{i+1}^{\dag} \right)
\end{equation}
}

\newcommand{\kldivergence}
{
	\begin{equation}\label{Fidelity}
		D_{KL} = S(\rho_{\text{measured}}, \rho_{\text{exponential}}) - S(\rho_{\text{measured}})
	\end{equation}
}

\newcommand{\fidelity}
{
	\begin{equation}\label{Fidelity}
	\frac{ S\left(P_\text{incoherent}, P_\text{expected}\right) 
									- S\left(P_\text{measured}, P_\text{expected}\right)}
						   { S\left(P_\text{incoherent}, P_\text{expected}\right) 
						   	- S\left(P_\text{expected}\right)} 
	\end{equation}
}

\newcommand{\correlations}
{
\begin{equation}\label{Correlations}
\left| \left<\hat{n}_i \hat{n}_j\right> - \left<\hat{n}_i\right> \left<\hat{n}_j\right> \right|
\end{equation}
}

\begin{document}

\title{A blueprint for demonstrating quantum supremacy with superconducting qubits}

\author{C. Neill$^{1}$}
\thanks{These authors contributed equally to this work.}
\author{P. Roushan$^{2}$}
\thanks{These authors contributed equally to this work.}
\author{K. Kechedzhi$^{3}$}
\author{S. Boixo$^{2}$}
\author{S. V. Isakov$^{2}$}
\author{V. Smelyanskiy$^{2}$}

\author{R. Barends$^{2}$}
\author{B. Burkett$^{2}$}
\author{Y. Chen$^{2}$}
\author{Z. Chen$^{1}$}
\author{B. Chiaro$^{1}$}
\author{A. Dunsworth$^{1}$}
\author{A. Fowler$^{2}$}
\author{B. Foxen$^{1}$}
\author{R. Graff$^{2}$}
\author{E. Jeffrey$^{2}$}
\author{J. Kelly$^{2}$}
\author{E. Lucero$^{2}$}
\author{A. Megrant$^{2}$}
\author{J. Mutus$^{2}$}
\author{M. Neeley$^{2}$}
\author{C. Quintana$^{1}$}
\author{D. Sank$^{2}$}
\author{A. Vainsencher$^{2}$}
\author{J. Wenner$^{1}$}
\author{T. C. White$^{2}$}
\author{H. Neven$^{2}$}
\author{J. M. Martinis$^{1,2}$}

\email{martinis@physics.ucsb.edu}

\affiliation{$^{1}$Department of Physics, University of California, Santa Barbara, CA 93106-9530, USA}
\affiliation{$^{2}$Google Inc., Santa Barbara, CA 93117, USA}
\affiliation{$^{3}$QuAIL, NASA Ames Research Center, Moffett Field, CA 94035, USA}

\maketitle

\textbf{
Fundamental questions in chemistry and physics may never be answered due to the exponential complexity of the underlying quantum phenomena.  
A desire to overcome this challenge has sparked a new industry of quantum technologies with the promise that engineered quantum systems can address these hard problems.
A key step towards demonstrating such a system will be performing a computation beyond the capabilities of any classical computer, achieving so-called quantum supremacy.
Here, using 9 superconducting qubits, we demonstrate an immediate path towards quantum supremacy.
By individually tuning the qubit parameters, we are able to generate thousands of unique Hamiltonian evolutions and probe the output probabilities.
The measured probabilities obey a universal distribution, consistent with uniformly sampling the full Hilbert-space.
As the number of qubits in the algorithm is varied, the system continues to explore the exponentially growing number of states.
Combining these large datasets with techniques from machine learning allows us to construct a model which accurately predicts the measured probabilities.
We demonstrate an application of these algorithms by systematically increasing the disorder and observing a transition from delocalized states to localized states.
By extending these results to a system of 50 qubits, we hope to address scientific questions that are beyond the capabilities of any classical computer.
}

\begin{figure}[b!]
	\includegraphics{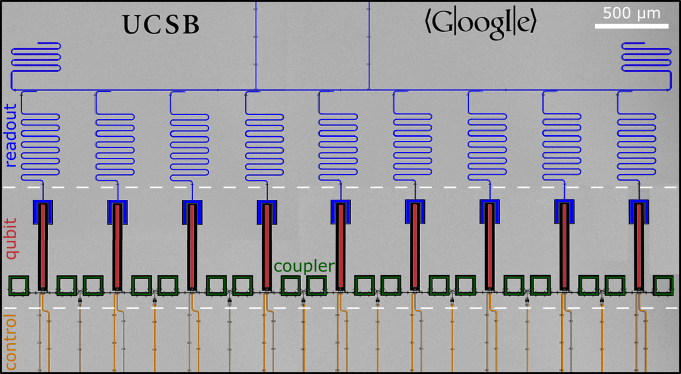} 
	\caption {
		\textbf{Device: nine-qubit array.}
		 Optical micrograph of the device.  Gray regions are aluminum, dark regions are where the aluminum has been etched away to define features.  Colors have been added to distinguish readout circuitry, qubits, couplers and control wiring.
	}
	\label{fig:device} 
\end{figure}

A programmable quantum system consisting of merely 50 to 100 qubits could revolutionize scientific research.
While such a platform is naturally suited to address problems in quantum chemistry and materials science \citep{feynman1982simulating, buluta2009quantum, lanyon2010towards, peruzzo2013variational}, applications range to fields as far as classical dynamics \citep{mezzacapo2015quantum} and computer science \citep{kadowaki1998quantum, boixo2014evidence, lloyd2013quantum, denchev2016computational}.
A key milestone on the path towards realizing these applications will be the demonstration of an algorithm which exceeds the capabilities of any classical computer - achieving quantum supremacy \citep{preskill_2012}. 
Sampling problems are an iconic example of algorithms designed specifically for this purpose \citep{aaronson2011computational,peropadre2016proposal,bremner2015average,boixo2016characterizing}. 
A successful demonstration of quantum supremacy would prove that engineered quantum systems, while still in their infancy, can outperform the most advanced classical computers.

Consider a system of coupled qubits whose dynamics uniformly explore all accessible states over time.
The complexity of simulating this evolution on a classical computer is easy to understand and quantify.
Since every state is equally important, it is not possible to simplify the problem, using a smaller truncated state-space.
The complexity is then simply given by asking how much classical memory does it take to store the state-vector.
Storing the state of a 46-qubit system takes nearly a petabyte of memory and is at the limit of the most powerful computers \citep{haner20170, boixo2016characterizing}.
Sampling from the output probabilities of such a system would therefore constitute a clear demonstration of quantum supremacy.
Note that this is only an upper bound on the number of qubits required - other constraints, such as computation time, may place practical limitations on even smaller system sizes.

Here, we experimentally illustrate a blueprint for demonstrating quantum supremacy.
We present data characterizing two basic ingredients required for any supremacy experiment: \textit{complexity} and \textit{fidelity}.
First, we demonstrate that the qubits can uniformly explore the Hilbert-space, providing a direct measure of algorithm complexity.
Next, we compare the measurement results with the expected behavior and show that the algorithm can be implemented with high fidelity.
Experiments for probing complexity and fidelity provide a foundation for demonstrating quantum supremacy.

The more control a quantum platform offers, the easier it is to embed diverse applications.
For this reason, we have developed superconducting gmon qubits with tunable frequencies and tunable interactions.
A photograph of the device used in this experiment is shown in Fig.\,\ref{fig:device}.
The device consists of three distinct sections:  control (bottom), qubits (center) and readout (top).
A detailed circuit diagram is provided in the supplementary material.

Each of our gmon qubits can be thought of as a nonlinear oscillator.
The Hamiltonian for the device is given by \hamiltonian where $\hat{n}$ is the number operator and $\hat{a}^{\dag}$ ($\hat{a}$) is the raising (lowering) operator.
The qubit frequency sets the coefficient $\delta_i$, the nonlinearity sets $\eta_i$ and the nearest neighbor coupling sets $g_i$.
The two lowest energy levels ($\ket{0}$ and $\ket{1}$) form the qubit subspace.
The higher energy levels of the qubits, while only virtually occupied, substantially modify the dynamics.

\begin{figure}[t!]
	\includegraphics{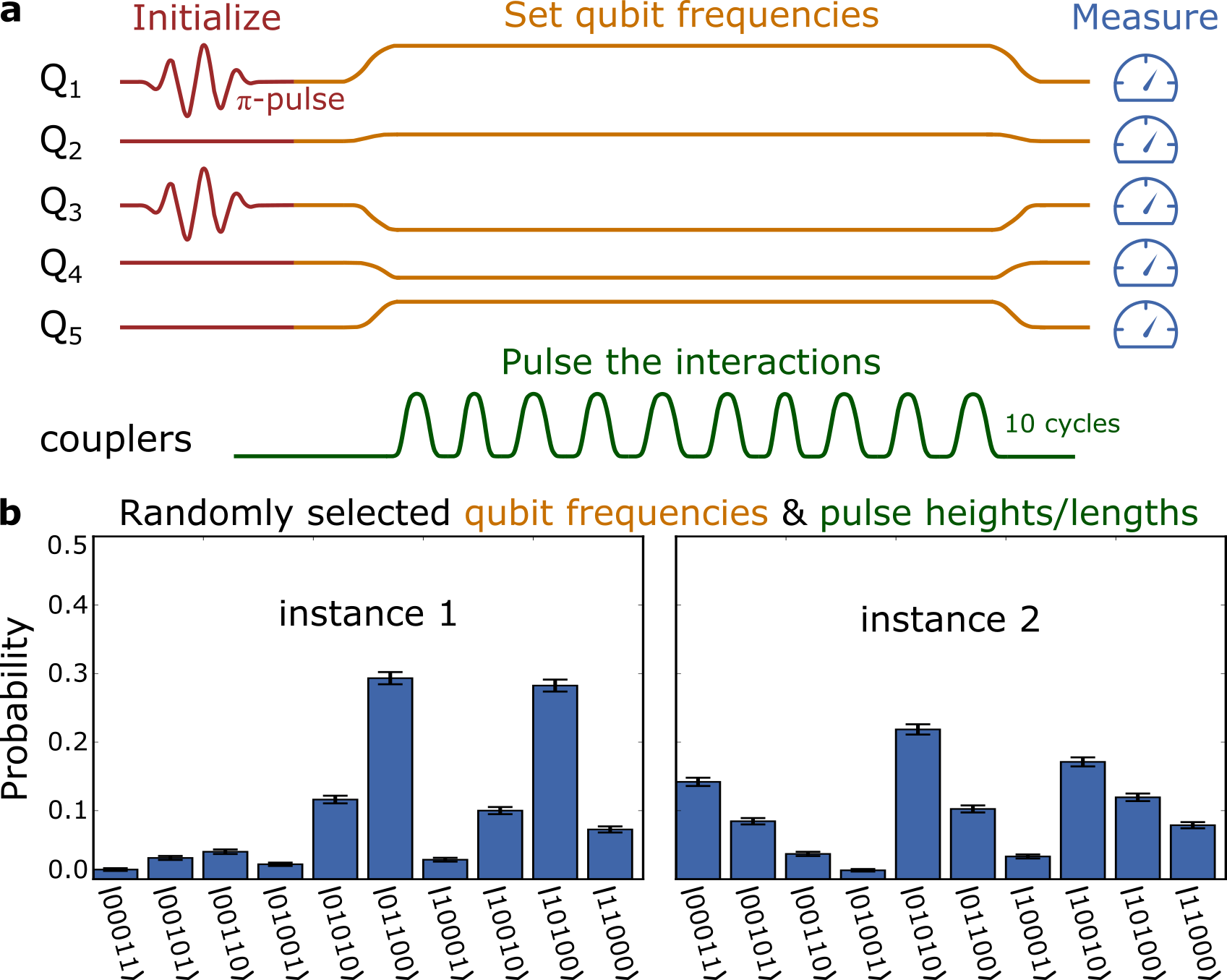} 
	\caption {
		\textbf{Protocol:  pulse sequence \& raw data.}
		\textbf{a} Five qubit example of the pulse sequences used in these experiments.  First, the qubits are initialized using microwave pulses (red).  Half of the qubits start in the ground state $\ket{0}$ and half start in the excited state $\ket{1}$. Next, the qubit frequencies are set using rectangular pulses (orange).  During this time, all the couplings are simultaneously pulsed (green); each pulse has a randomly selected duration.  Lastly, we measure the state of every qubit. The measurement is repeated many times in order to estimate the probability of each output state.
		\textbf{b} We repeat this pulse sequence for randomly selected control parameters.  For each instance, the qubit frequencies, coupling pulse heights and lengths are varied.  Here, we plot the measured probabilities for two instances after 10 coupler pulses (cycles).  Error bars ($\pm 3$ standard deviations) represent the statistical uncertainty from 50,000 samples.
	}
	\label{fig:rawdata} 
\end{figure}

In Fig.\,\ref{fig:rawdata} we outline the experimental procedure and provide two instances of the raw output data.
Panel $\textbf{a}$ shows a five-qubit example of the pulses used to control the qubits.
First, the system is initialized (red) by placing half of the qubits in the excited state, e.g. $\ket{00101}$.
The dynamics result from fixing the qubit frequencies (orange) and simultaneously ramping all of the nearest-neighbor interactions on and then off (green).
The shape of the coupling pulse is chosen to minimize leakage out of the qubit subspace \citep{martinis2014fast}.
After the evolution, we simultaneously measure the state of every qubit.
Each measurement results in a single output state, such as $\ket{10010}$; the experiment is repeated many times in order to estimate the probability of every possible output state.
We then carry out this procedure for randomly chosen values of the qubit frequencies, the coupler pulse lengths, and the coupler pulse heights. 
The probabilities of the various output states are shown in panel $\textbf{b}$ for two instances of the evolution after 10 coupler pulses (cycles).
The height of each bar represent the probability with which that output state appeared in the experiments.

It is important to note that the Hamiltonian in Eq.\,1 conserves the total number of excitations.
This means that because we start in a state with half the qubits excited, we should also end in a state with half the qubits excited.
However, most experimental errors do not obey this symmetry, allowing us to identify and remove erroneous outcomes.
While this helps to reduce the impact of errors, it slightly reduces the size of the Hilbert-space.
For N qubits, the number of states is given by the permutations of N/2 excitations in N qubits and is approximately $2^N / \sqrt{N}$.
As an example, a 64 qubit system would access roughly $2^{61}$ states under our protocol.

The measured probabilities, while they appear largely random, provide significant insight into the quantum dynamics.
A key feature of these datasets are the rare, taller-than-average peaks - analogous to the high intensity regions of a laser's speckle pattern.
These highly-likely states serve as a fingerprint of the underlying evolution and provide a means for verifying that the desired evolution was properly generated.
The distribution of these probabilities provides evidence that the dynamics coherently and uniformly explore the Hilbert-space.

\begin{figure}[t!]
	\includegraphics{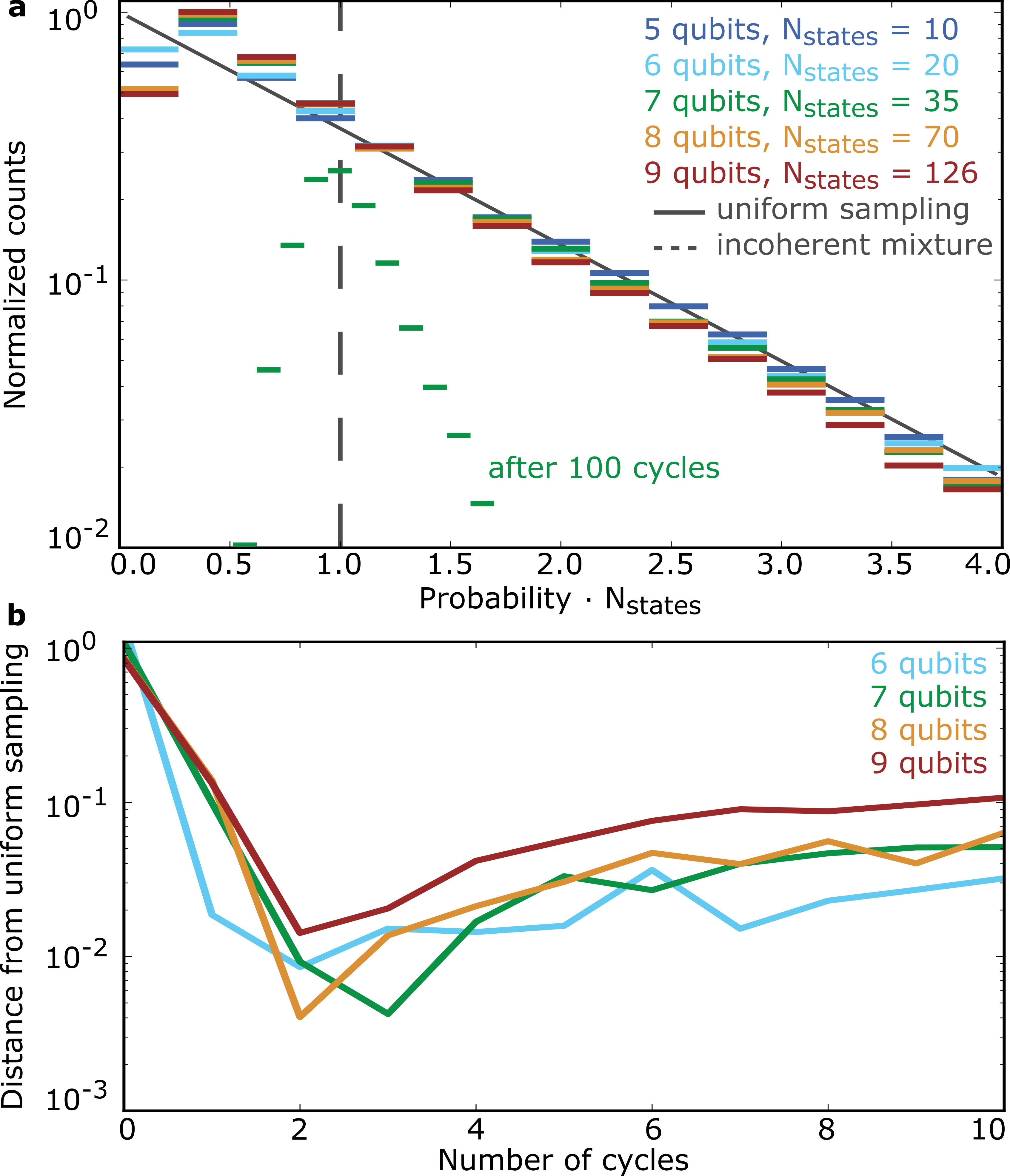} 
	\caption {
		\textbf{Complexity: uniform sampling of an exponentially growing state-space.}
		\textbf{a}  Histograms of the raw probabilities (see Fig.\,1b) for 5 to 9 qubit experiments, after five cycles of evolution.  Before making the histogram, probabilities are weighted by the number of states in the Hilbert-space, this places all of the curves onto a universal axis.  The width of the bars represents the size of the bins used to construct the histogram.  The data is taken from over $29.7$ million experiments.  For dynamics which uniformly explore all states, this histogram decays exponentially; an exponential decay is shown as a solid line for comparison.  For contrast, we plot a histogram of the probabilities for 7 qubits after 100 cycles.  Here, decoherence dominates and we observe a tall narrow peak around 1.
		\textbf{b}  In order to measure convergence of the measured histogram to an exponential distribution, we compute their distance as a function of the number of cycles.  Distance is measured using the KL-divergence (see Eq.\,2).  We find that a maximum overlap occurs after just two cycles, following which decoherence increases their distance.
	}
	\label{fig:histograms} 
\end{figure}

In Fig.\,\ref{fig:histograms} we use the measured probabilities to show that the dynamics uniformly explore the Hilbert-space for experiments ranging from 5 to 9 qubits.
We begin by measuring the output probabilities after 5 cycles for between 500 and 5000 unique instances.
In order to compare experiments with different number of qubits, the probabilities are weighted by the number of states in the Hilbert-space.
Fig.\,\ref{fig:histograms}a shows a histogram of the weighted probabilities where we find nearly universal behavior.
Small probabilities (less than $1/\text{N}_{\text{states}}$) appear most often and probabilities as large as $4/\text{N}_{\text{states}}$ show up with a frequency of around 1\%.
In stark contrast to this, we observe a tall narrow peak centered around 1 for longer evolutions whose duration is comparable to the coherence time of the qubits.

A quantum system which uniformly explores all states is expected to have an exponential distribution of weighted probabilities.
The dark solid line in Fig.\,\ref{fig:histograms}a corresponds to such a distribution and is simply given by $e^{-\text{probability} \times \text{N}_{\text{states}}}$; this is also referred to as a Porter-Thomas distribution \citep{boixo2016characterizing, porter1956fluctuations}.
The universal and exponential behavior of the data leads us to conclude that the dynamics are uniformly exploring the state-space.
Deviations from an exponential distribution are the result of decoherence which drives the output states to appear with equal probability; this is the behavior that we observe at long times.
This demonstration of dynamics that take advantage of the full exponentially growing number of states (a direct probe of computational complexity) is a key ingredient for experimentally demonstrating quantum supremacy.

In Fig.\,\ref{fig:histograms}b we study the number of cycles it takes for the system to uniformly explore all states by comparing the measured probabilities to an exponential distribution.
After each cycle, we compare the measured histogram to an exponential decay.
The distance between these two distributions is measured using the KL-divergence $D_{KL}$, \kldivergence
where the first term is the cross-entropy between the measured distribution $\rho_{\text{measured}}$ and an exponential distribution $\rho_{\text{exponential}}$, and the second term is the self-entropy of the measured distribution.
The entropy of a set of probabilities is given by $S\left(P\right) = -\sum_i p_i \log\left(p_i\right)$ and the cross-entropy of two sets of probabilities is given by $S\left(P, Q\right) = -\sum_i p_i \log\left(q_i\right)$.
Their difference, the KL-divergence, is zero if and only if the two distributions are equivalent.

We find that the experimental probabilities closely resemble an exponential distribution after just two cycles.
For longer evolutions, decoherence reduces this overlap.
These results suggests that we can generate very complex dynamics with only two pulses - a surprisingly small number.

\begin{figure}[t!]
	\includegraphics{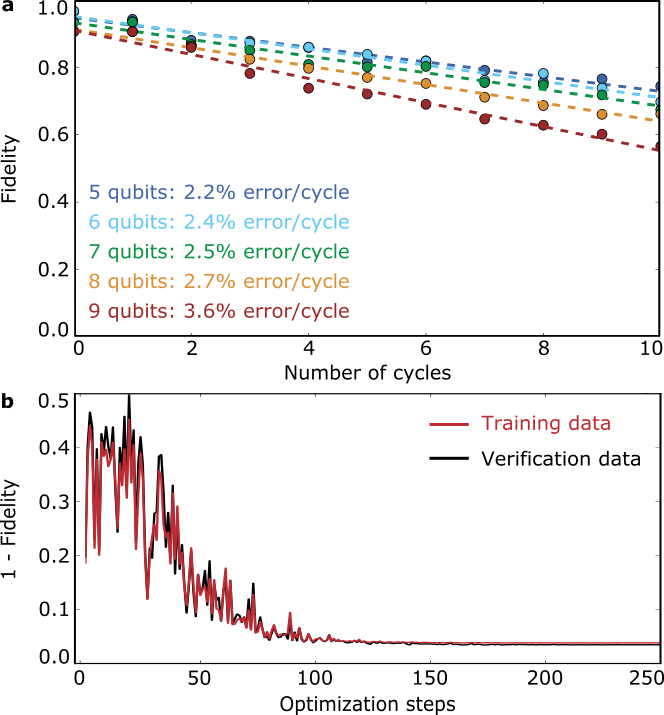} 
	\caption {
		\textbf{Fidelity:  learning a better control model.} 
		\textbf{a} Average fidelity decay versus number of cycles for 5 to 9 qubit experiments (circles).  The fidelity is computed using Eq.\,3.  The error per cycle, presented inset, is the slope of the dashed-line which best fits the data.
		\textbf{b} Using the fidelity as a cost-function, we learn optimal parameters for our control model.  Here, we take half of the experimental data and use this to train our model.  The other half of the data is used to verify this new model; the optimizer does not have access to this data.  The corresponding improvement in fidelity of the verification set provides evidence that we are indeed learning a better control model.
	}
	\label{fig:fidelity} 
\end{figure}

In addition to demonstrating an exponential scaling of complexity, it is necessary to characterize the algorithm fidelity.
Determining the fidelity requires a means for comparing the measured probabilities $P_\text{measured}$ with the probabilities expected from the desired evolution $P_\text{expected}$.
Based on the proposal outlined in Ref.\,\citep{boixo2016characterizing}, we use the cross-entropy to quantify the fidelity \fidelity
where $P_\text{incoherent}$ stands for an incoherent mixture with each output state given equal likelihood - this is the behavior that we observe after many cycles.
When the distances between the measured and expected probabilities is small, the fidelity approaches 1.
When the measured probabilities approach an incoherent mixture, the fidelity approaches 0.

\begin{figure}[t!]
	\includegraphics{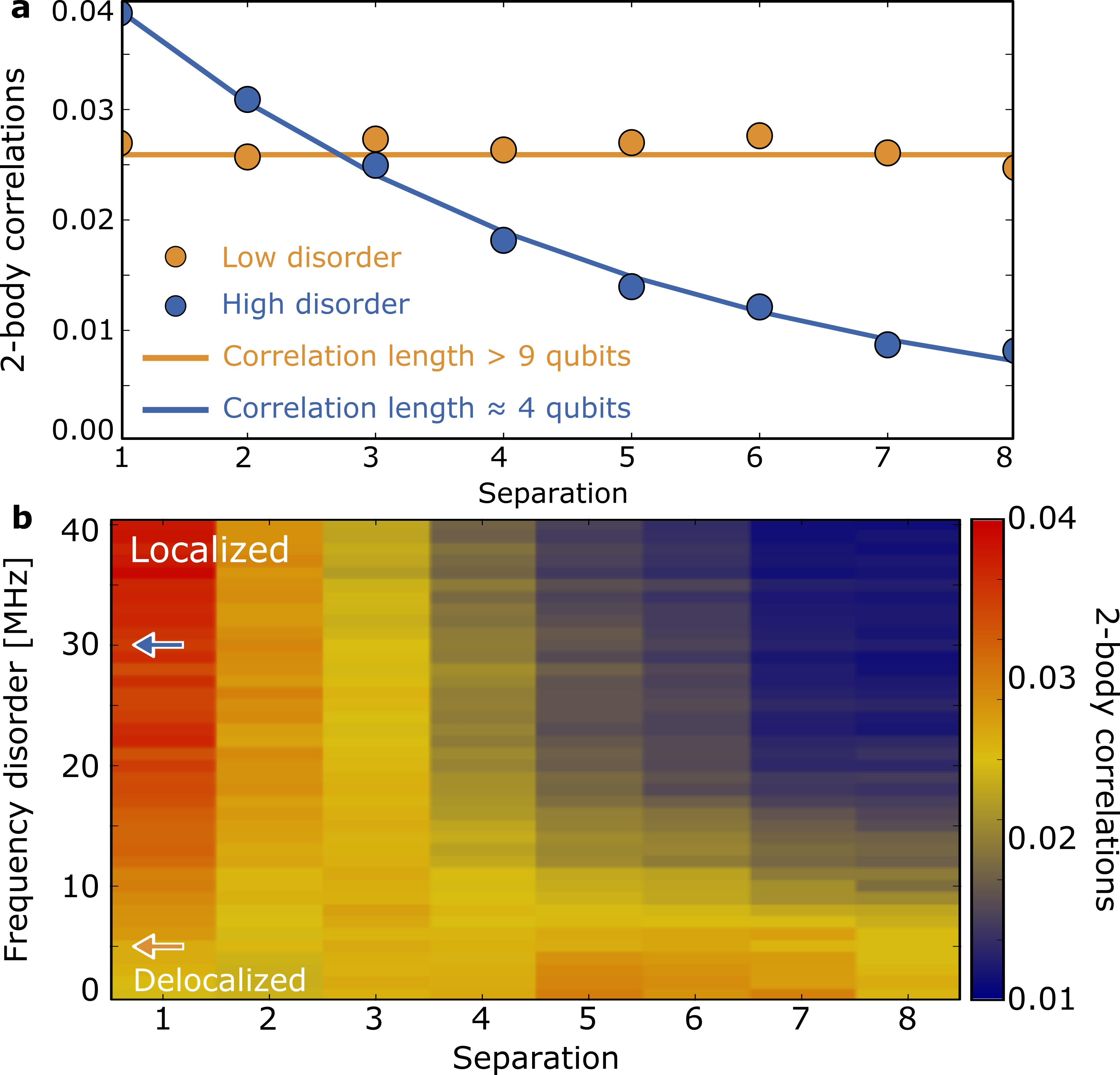} 
	\caption {
		\textbf{Applications:  localization \& delocalization.}
		\textbf{a} Average 2-body correlations (see Eq.\,4) as a function of the separation between qubits.  The data is shown for two values of disorder strength.  At low disorder (gold) the qubit frequencies are set over a range of $\pm$\,5\,MHz and the 2-body correlations are independent of separation (i.e. qubits at the ends of the chain are just as correlated as nearest neighbors). At high disorder (blue) the qubit frequencies are set over a range of $\pm$\,30\,MHz and we find an exponential decay in correlations as a function of separation.
		\textbf{b} We map out the correlations as a continuous function of the frequency disorder.  Arrows indicate the location of line cuts used in panel \textbf{a}.  We observe a clear transition from long-range to short-range correlations.
	}
	\label{fig:localization} 
\end{figure}

In Fig.\,\ref{fig:fidelity}a we show that the desired evolution can be implemented with high fidelity.
We find that at short times the fidelity decays linearly with increasing number of cycles, fits to the data are shown as dashed lines.
The slope of these lines measures the error per cycle; this slope is displayed inset for each number of qubits.
We find that the error scales with the number of qubits at a rate of around 0.4\% error/qubit/cycle.
If such an error rate extends to larger systems, we will be able to perform 60 qubit experiments of depth 2 while still remaining above 50\% total error.
These results provide promising evidence that quantum supremacy may be achievable using existing technology.

Predicting the expected probabilities is a major challenge.  
First, substantial effort has been taken to accurately map the control currents to Hamiltonian parameters; the detailed procedure for constructing this map is outlined in the supplementary materials.  
Second, we model the Hamiltonian using only single qubit calibrations, which we find is accurate even when all the couplers are used simultaneously.  
This is a scalable approach to calibration.
Third, when truncating the Hamiltonian to qubits with two-levels, we find poor agreement both theoretically and experimentally.  
We find a 3-level description must be used to account for virtual transitions to the second excited state during the evolution.  
When including these states, truncating to a fixed number of excitations lowers the size of the computational Hilbert space from $3^N$ to approximately $0.15 \times 2.42^N$ (see Table\,1 in supplement): thus a nine-qubit experiment requires accurately modeling a 414-dimensional unitary operation.  
Determining how many of these states are needed for sufficient accuracy depends on the magnitude of the coupling and is an open question, but should scale somewhere between $2.0^N$ and $2.5^N$. 

In Fig.\,\ref{fig:fidelity}b we show how techniques from machine learning were used to achieve low error rates.
In order to set the matrix-elements of the Hamiltonian, we build a physical model for our gmon qubits.
This model is parameterized in terms of capacitances, inductances and control currents.
The parameters in this model are calibrated using simple single qubit experiments (see supplement).
Here, we use a search algorithm to find offsets in the control model which minimize the error (1 - Fidelity).
Figure\,\ref{fig:fidelity}b shows the error, averaged over cycles, versus the number of optimization steps.
Prior to training the model, the data is split into two halves - a training set (red) and a verification set (black).
The optimization algorithm is only provided access to the training set, the verification set is used only to verify the optimal parameters.

We find that the error in both the training set and the verification set fall significantly by the end of the optimization procedure.
The high degree of correlation between the training and verification data suggests that we are genuinely learning a better physical model. 
Optimizing over more parameters does not further reduce the error.
This suggests that the remaining error is not control but results from decoherence.
Using the cross-entropy as a cost function for optimizing the parameters of a physical model was the key to achieving high-fidelity control in this experiment.

Ideally, in addition to exponential complexity and high fidelity, a quantum platform should offer valuable applications.
In Fig.\,\ref{fig:localization} we consider applications of our algorithms to many-body physics where the exponential growth in complexity is a significant barrier to ongoing research \cite{schreiber2015observation, basko2006metal, pal2010many, santos2011entropy, rigol2008thermalization, polkovnikov2011colloquium}.
By varying the amount of disorder in the system, we are able to study disorder-induced localization.
This is done using 2-body correlations \correlations which we average over qubit-pairs, cycles and instances.
In panel \textbf{a}, we plot the average 2-body correlations versus the separation between qubits.
This experiment is performed for both low and high disorder in the qubit frequencies, shown in gold and blue respectively.
In panel \textbf{b}, this experiment is carried out as we continuously vary the amount of disorder.

At low disorder, we find that the correlations are independent of separation - qubits at opposite ends of the chain are as correlated as nearest neighbors.
At high disorder, the correlations fall off exponentially with separation.  
The rate at which this exponential decays allows us to determine the correlation length.
A fit to the data is shown in Fig.\,\ref{fig:localization}a as a solid blue line where we find a correlation length of roughly 4 qubits.
The study of localization and delocalization in interacting systems provides a promising application of our algorithms.

Here, we have demonstrated an immediate path towards quantum supremacy.
We show that the algorithm complexity scales exponentially with the number of qubits and can be implemented with high fidelity.
If similar error rates are achievable in future devices with around 50 qubits, we will be able to explore quantum dynamics that are inaccessible otherwise.

\textbf{Acknowledgments:} This work was supported by Google.  C.Q. and Z.C. acknowledge support from the National Science Foundation Graduate Research Fellowship under Grant No. DGE-1144085.  Devices were made at the UC Santa Barbara Nanofabrication Facility, a part of the NSF funded National Nanotechnology Infrastructure Network.

\textbf{Author Contributions:} C.N. designed and fabricated the device. C.N. and P.R. designed the experiment. C.N. performed the experiment and analyzed the data. C.N.,  K.K. and V.S. developed the physical control model.  S.B. and S.I. numerically validated the protocol for large qubit arrays. All members of the UCSB and Google team contributed to the experimental setup and to the manuscript preparation.

\bibliographystyle{unsrtnat}

\end{document}

\title{Supplementary Information for "A blueprint for demonstrating quantum supremacy with superconducting qubits"}

\author{C. Neill$^{1}$}
\thanks{These authors contributed equally to this work.}
\author{P. Roushan$^{2}$}
\thanks{These authors contributed equally to this work.}
\author{K. Kechedzhi$^{3}$}
\author{S. Boixo$^{2}$}
\author{S. V. Isakov$^{2}$}
\author{V. Smelyanskiy$^{2}$}

\author{R. Barends$^{2}$}
\author{B. Burkett$^{2}$}
\author{Y. Chen$^{2}$}
\author{Z. Chen$^{1}$}
\author{B. Chiaro$^{1}$}
\author{A. Dunsworth$^{1}$}
\author{A. Fowler$^{2}$}
\author{B. Foxen$^{1}$}
\author{R. Graff$^{2}$}
\author{E. Jeffrey$^{2}$}
\author{J. Kelly$^{2}$}
\author{E. Lucero$^{2}$}
\author{A. Megrant$^{2}$}
\author{J. Mutus$^{2}$}
\author{M. Neeley$^{2}$}
\author{C. Quintana$^{1}$}
\author{D. Sank$^{2}$}
\author{A. Vainsencher$^{2}$}
\author{J. Wenner$^{1}$}
\author{T. C. White$^{2}$}
\author{H. Neven$^{2}$}
\author{J. M. Martinis$^{1,2}$}

\email{martinis@physics.ucsb.edu}

\affiliation{$^{1}$Department of Physics, University of California, Santa Barbara, CA 93106-9530, USA}
\affiliation{$^{2}$Google Inc., Santa Barbara, CA 93117, USA}
\affiliation{$^{3}$QuAIL, NASA Ames Research Center, Moffett Field, CA 94035, USA}

\maketitle

\section{{Qubit architecture}}

\begin{figure}[b]
	\centering
	\includegraphics{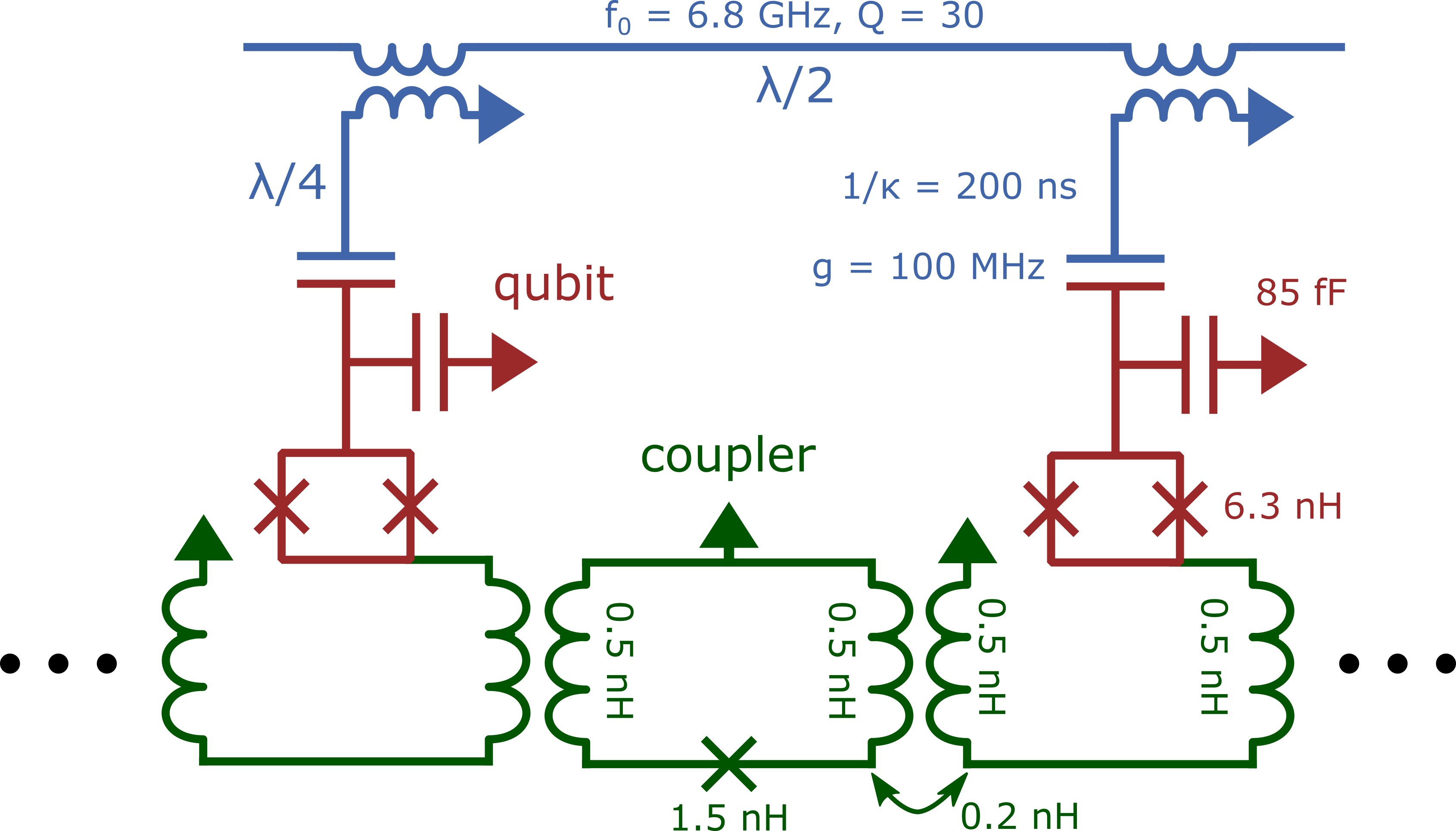}
	\caption 
	{
		\textbf{Device.}
		 Circuit diagram for the device shown in Fig.\,1 of the main text.
		 The qubit SQUID and coupler junction inductances are the effective linear values at zero phase.
	}
	\label{fig:device}
\end{figure}

The more control a quantum platform offers, the simpler it is to embed diverse applications.
For this reason, we have developed superconducting gmon qubits with tunable frequencies and tunable interactions.
An optical micrograph of the device is shown in Fig.\,1 of the main text and the corresponding circuit diagram is shown in Fig.\,\ref{fig:device}.
The qubits consist of a capacitor (coplanar-waveguide strip), a flux-tunable DC SQUID and a shunt inductor.
While the qubits are similar to typical planar transmons (unbiased frequency of 6.2\,GHz, non-linearity of 180\,MHz), the shunt inductor provides coupling to an RF SQUID.
Flux into this RF SQUID allows for tunable coupling between neighboring qubits.
The voltage divider created by the DC SQUID and shunt inductor protect the qubit from capacitive losses in the coupling circuit.
Each qubit is dispersively coupled to a readout resonator which are themselves coupled to a low-Q Purcell filter.
Twenty-six control lines are used to drive microwave rotations (coupling of 50\,aF), set the qubit frequencies (mutual of 1\,pH), and bias the couplers (mutual of 1\,pH).
Interconnects and crossovers are formed using aluminum air-bridges.
The qubits have energy relaxation times $T_1 \approx 10-15\,\mu$s and ramsey dephasing times of around $5\,\mu$s near the flux-insensitive point.

\section{{Raw data with predictions}}

\begin{figure}[H]
	\centering
	\includegraphics{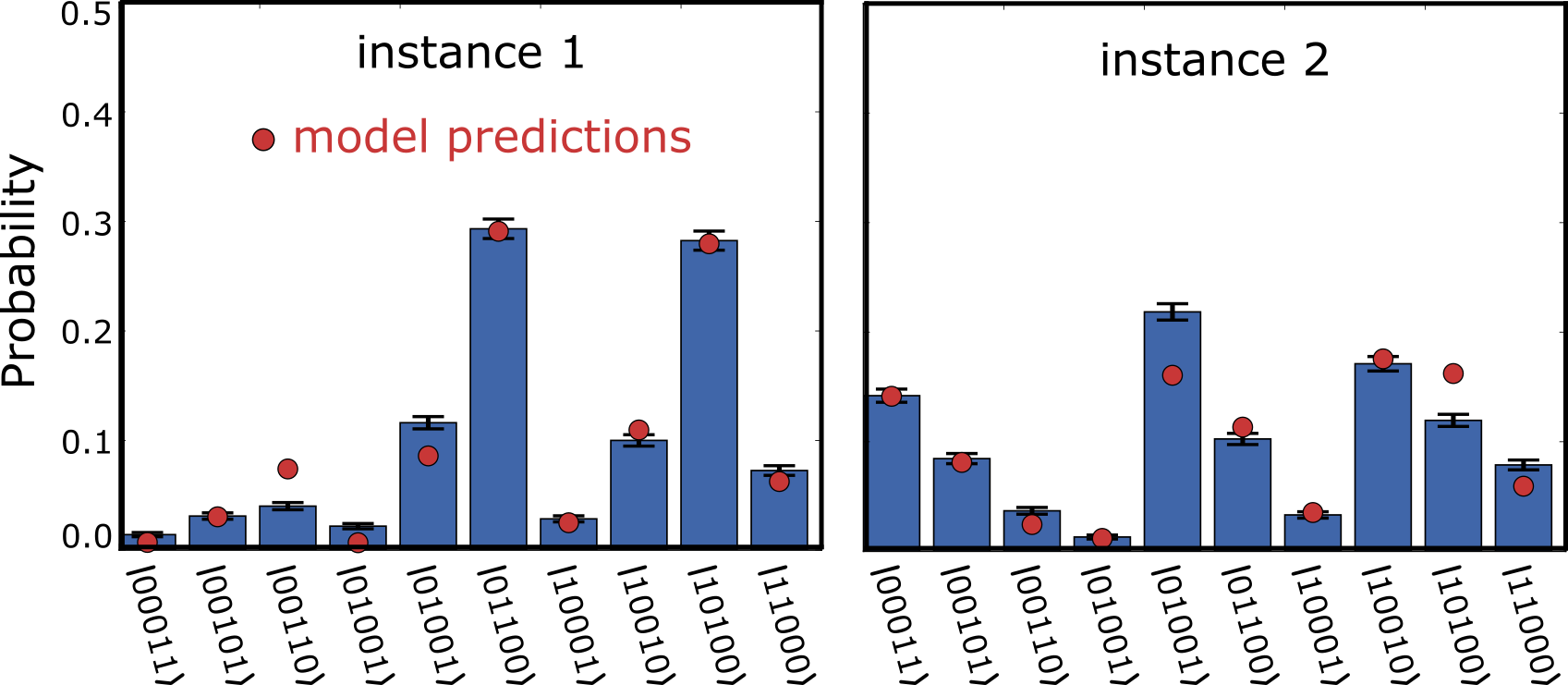}
	\caption 
	{
		\textbf{Raw data with model predictions.}
		Measured probabilities for two instances of a 5 qubit experiment after 10 cycles (blue bars).
		Error bars (3 standard deviations) represent the statistical error from 50,000 samples.
		This is the same data used in Fig.\,1 of the main-text, however, now we overlay the expected probabilities (red circles).
	}
	\label{fig:rawdata}
\end{figure}

In the main-text, the cross-entropy is used to show that we are able to accurately predict the measured probabilities.
Here, we show the measured probabilities with the predictions overlaid onto the data (Fig.\,\ref{fig:rawdata}).
We find a strong resemblance between the data and the predictions.
While less quantitative than Fig.\,4a in the main-text, this provides an intuitive and visual demonstration that we have developed an accurate control model.

\clearpage
\section{{Histogram of probabilities without normalization}}

\begin{figure}[H]
	\centering
	\includegraphics{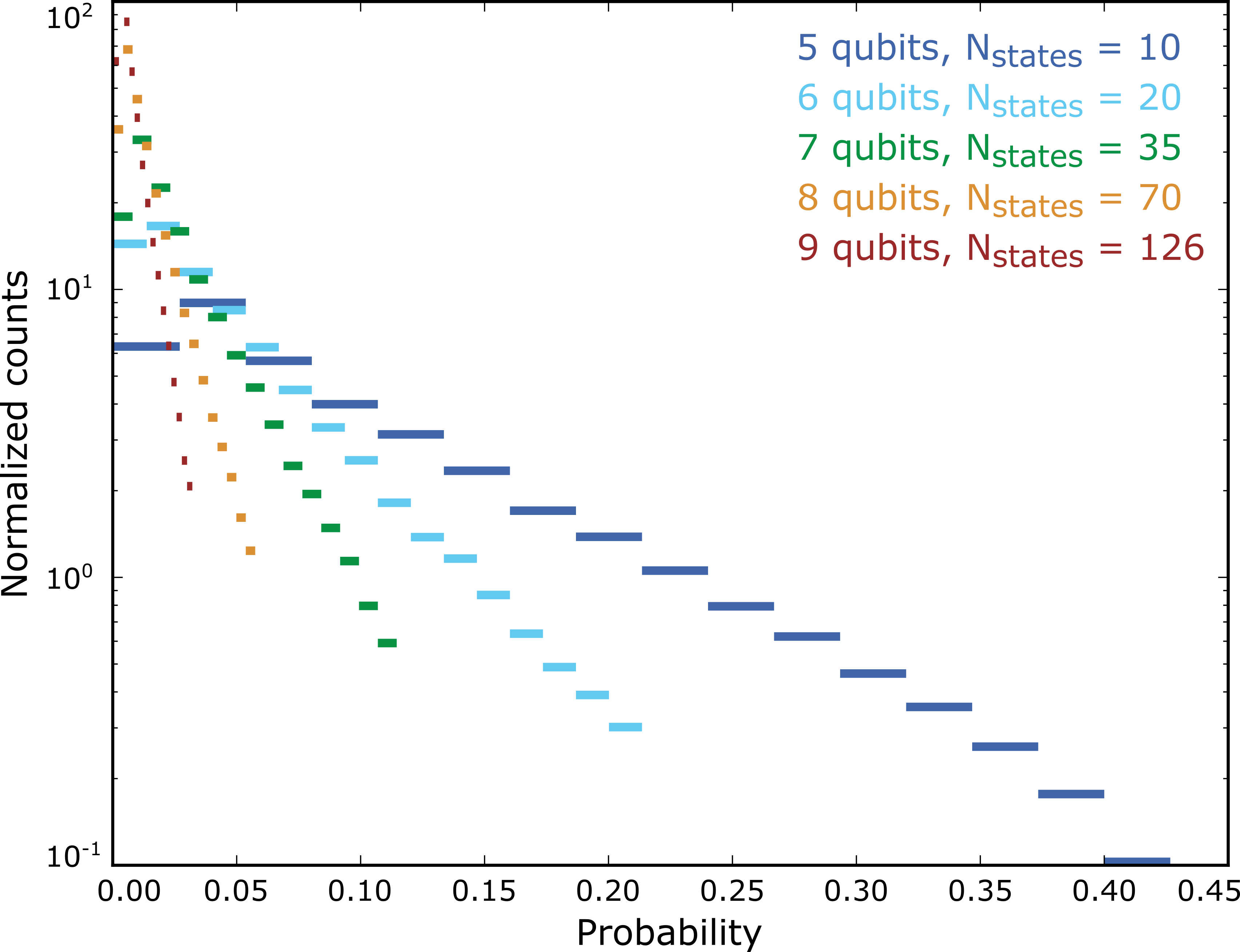}
	\caption 
	{
		\textbf{Histogram of probabilities.}
		Histograms of the raw output probabilities for 5 to 9 qubit experiments after 5 cycles of evolution.
		The width of the bars represent the size of the bins used to construct the histogram.
		This is the same data used in Fig.\,2 of the main-text, however, here we do not normalize the probabilities by the number of states.
	}
	\label{fig:histograms}
\end{figure}

The output probabilities of interacting quantum systems (after sufficiently long evolution) all obey a single universal distribution, known as a Porter-Thomas distribution.
The distribution of probabilities p has the form $e^{-p \times N_\text{states}}$ where $N_\text{states}$ is the dimension of the state space.
In the main text, we plot histograms of the measured probabilities weighted by $N_\text{states}$ - this places the data onto a universal axis.
For completeness, we present histograms of the same datasets without weights (see Fig.\,\ref{fig:histograms}).

\section{{Hilbert-space dimension}}

Determining how many states are required to accurately reproduce the experimental data is key to understanding the complexity of our algorithms.
In the main text, we show that the system uniformly explores a Hilbert space that grows as $2^N / \sqrt{N}$.
Therefore, this expression provides a lower bound on the complexity.
However, higher states of each qubit (e.g. $\ket{2}$) do modify the dynamics and including these states in the simulations increases the computational complexity (see Appendix for more details).

In Table\,1 we consider the number of states needed under various truncations schemes.
The first column is the number of qubits ranging from 3 to 9.
The second column is the number of states typically associated with qubit simulations ($2^N$) and is shown for comparison.
The third column is the number of states if each qubit is treated as a 2-level system and truncated to a subspace with half the qubits excited - this scales roughly as $2^N / \sqrt{N}$.
This is the lower bound on the scaling of complexity as the system uniformly explores these states (see main text).
In the fourth column we consider all states where only a single qubit is in the state $\ket{2}$ - we refer to these states as single doublons.
An example of such a state for 8 qubits and 4 excitation would be $\ket{02100100}$.
States of this form are the closest in energy to the qubit-subspace and, consequently, most significantly modify the evolution.
In the last column we consider qutrits that have been truncated to the subspace with the correct number of excitations and includes multi-doublon states.

\begin{table}[H]
	\begin{center}
		\begin{tabular}{ || M{0.5cm} | M{1.5cm}| M{1.5cm} | M{1.5cm}| M{1.75cm} || } 
			\hline
			N & 2-Levels & 2-Levels truncated & Single doublon & 3-Levels truncated \\
			\hline\hline
			3 & 8 & 3 & 6 & 6 \\ 
			\hline
			4 & 16 & 6 & 10 & 10  \\
			\hline
			5 & 32 & 10 & 15 & 15  \\
			\hline
			6 & 64 & 20 & 50 & 50  \\
			\hline
			7 & 128 & 35 & 77 & 77  \\
			\hline
			8 & 256 & 70 & 238 & 266  \\
			\hline		
			9 & 512 & 126 & 378 & 414  \\
			\hline
		\end{tabular}
		\caption{Hilbert-space dimension versus number of qubits $N$ for various truncation schemes.  The second column (2-levels) is simply $2^N$.  The next column is the number of states after truncating to a subspace with a fixed number of excitations.  The third column (single doublon) includes additional states where a single qubit is in the 2-state (e.g. $\ket{02100100}$).  The final column treats each qubit as a 3-level system and truncates to a fixed number of excitations.  This includes all multi-doublon states such as $\ket{02020000}$.}
	\end{center}
	\label{tab:complexity}
\end{table}

\begin{figure}[H]
	\centering
	\includegraphics{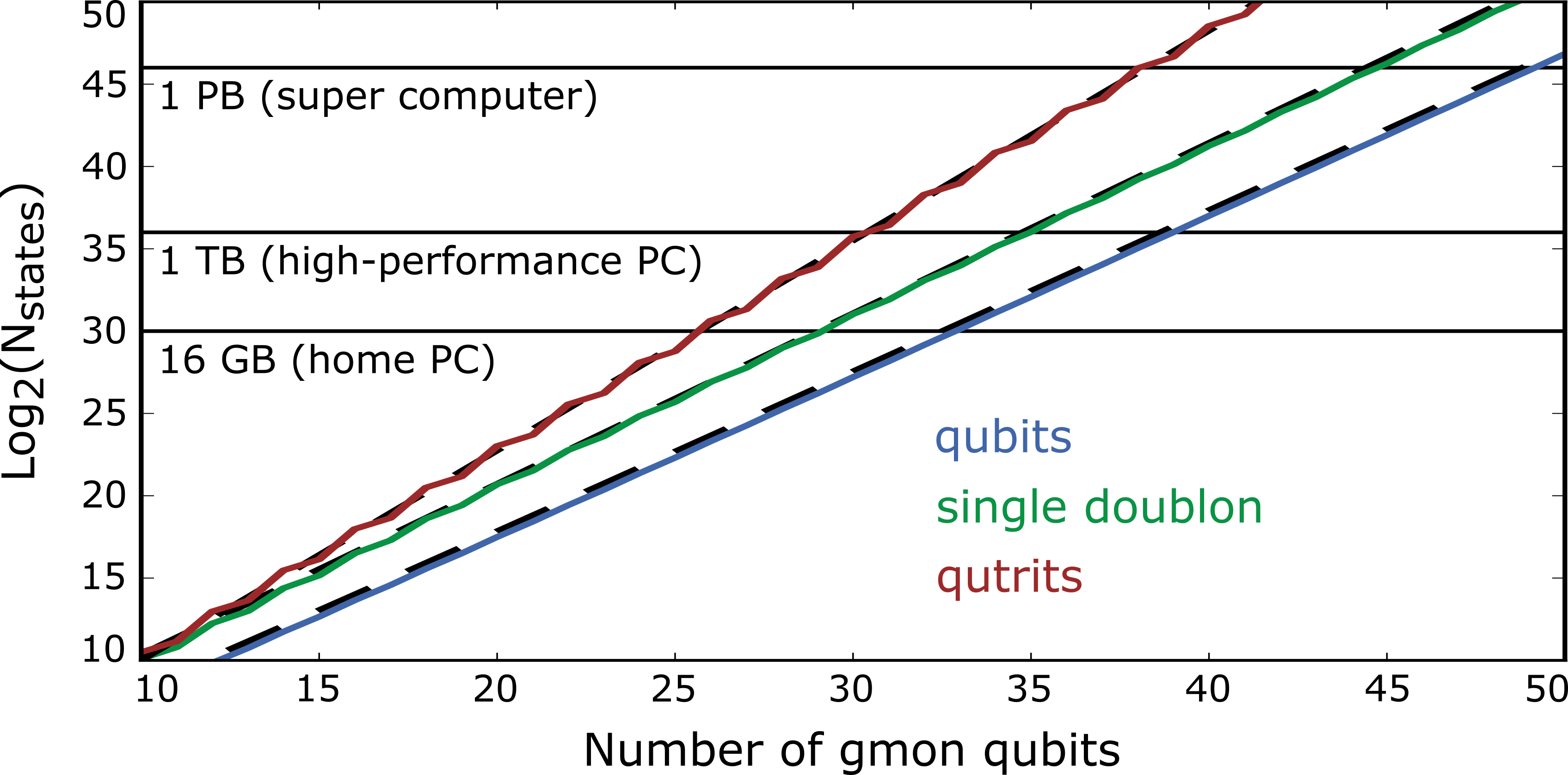}
	\caption 
	{
		\textbf{Complexity scaling.}
		Hilbert-space dimension versus number of qubits for 10 to 50 qubits for various truncation schemes.
		In every case, we only consider states with a fixed number of excitations.
		The exact scaling for qubits, single doublons and qutrits are shown in blue, green and red respectively.
		The corresponding dashed lines are approximate scalings.
		For qubits, this is given by $2^N/\sqrt{N}$.
		For the single-doublon subspace, we find that $2.05^N$ is an accurate approximation.
		For qutrits, we find that $0.15 \times 2.42^N$ is an accurate approximation.
		Horizontal lines correspond to memory requirements on a classical computer assuming the state is represented using complex 64-bit numbers.
	}
	\label{fig:complexity}
\end{figure}

In Fig.\,\ref{fig:complexity} we plot the number of states for each of these truncation schemes for 10 to 50 gmon qubits as solid colored lines.
We have taken $\log_2$ of the vertical axis so that it represents an effective number of qubits.
In addition to the exact scaling, we plot approximate expressions as dashed black lines.
We find that $2^N/\sqrt{N}$ is an accurate scaling for two-level systems assuming fixed number of excitations.
Fitting the results for the single-doublon and qutrit subspaces, we find approximate scalings of $2.05^N$ and $0.15 \times 2.42^N$.
For comparison, we plot the classical memory requirements as horizontal lines for a typical 16 GB home computer, a 1 TB high-performance computer, and a 1 PB super computer.
A home computer may be able to simulate 33, 28 or 26 qubits depending on the truncation scheme.
A super computer would be limited to 47, 44, or 37 qubits - this is the the number of qubits required to achieve quantum supremacy.
For further details on the complexity of these algorithms see Appendix.

\section{{Post-selection}}

\begin{figure}[H]
	\centering
	\includegraphics{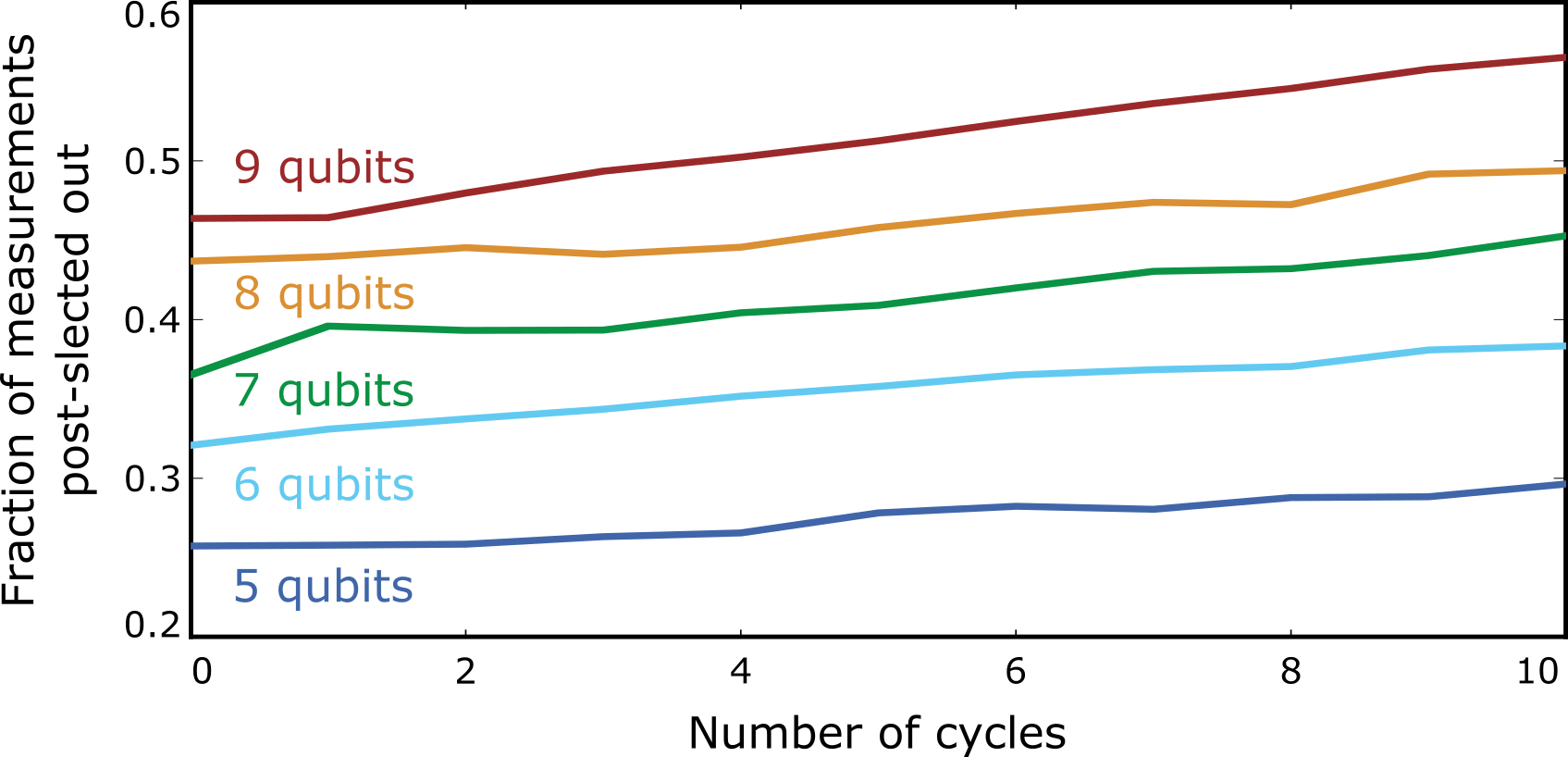}
	\caption 
	{
		\textbf{Post-selection.}
		The fraction of the measurements that were rejected during post-selection is plotted versus the number of cycles for 5 to 9 qubit experiments.
	}
	\label{fig:postselection}
\end{figure}

The control sequences used in this experiment are designed to conserve the total number of excitations.
However, experimental imperfections, such as measurement error and photon loss, can change the number of excitations.
The ability to identify and remove erroneous outcomes, while advantageous, comes at the expense of reduced data rates.
In Fig.\,\ref{fig:postselection} we plot the fraction of measurements that were thrown away as a function of the number of cycles for 5 to 9 qubit experiments.
Initially (at 0 cycles), the number of erroneous outcomes scales as 5.0\%/qubit, consistent with measurement infidelity.
The number of erroneous outcomes scales linearly with the number of cycles at a rate of 0.1\%/qubit/cycle, consistent with photon loss.

\section{{Calibrations:  Overview}}

Arbitrary time-dependent control of a well-specified Hamiltonian is a staggering control problem.
In the next four sections, we describe the calibration procedures used to tackle this problem.
While there is still a great deal of work left before completely arbitrary control is feasible, this work represents a significant step in that direction.

Converting time-dependent pulses into time-dependent matrix elements is done in two steps.
First, the signals produced at room temperature are not necessarily what reach the target device (qubit or coupler SQUID).
Nonidealities include pulse-distortion, relative timings, and crosstalk (see sections A, B \& C).
Second, a physical model of the device is required to convert the signals at the target device to matrix-elements of a Hamiltonian (see section D).

\subsection{{Calibration 1:  Pulse distortion}}

We observe that our control pulses undergo frequency-dependent attenuation.
The transfer-function between room temperature and the target-device is modeled as
\begin{equation}
\text{H}\left(\omega\right) = 1 + \epsilon \frac{i \omega \tau}{1 + i \omega \tau}
\end{equation}
where $\epsilon$ is the fraction of the pulse height that undergoes distortion (typically $\epsilon \approx 1$\%) and $\tau$ is a characteristic time-scale (typical values are 10\,ns and 70\,ns).
This expression can be derived using a simple model, treating the CPW bias-line as an inductor network with lossy asymmetric return paths.
While the physical origin of the pulse-distortion is not completely understood, this model is consistent with the data.

\begin{figure}[H]
	\centering
	\includegraphics{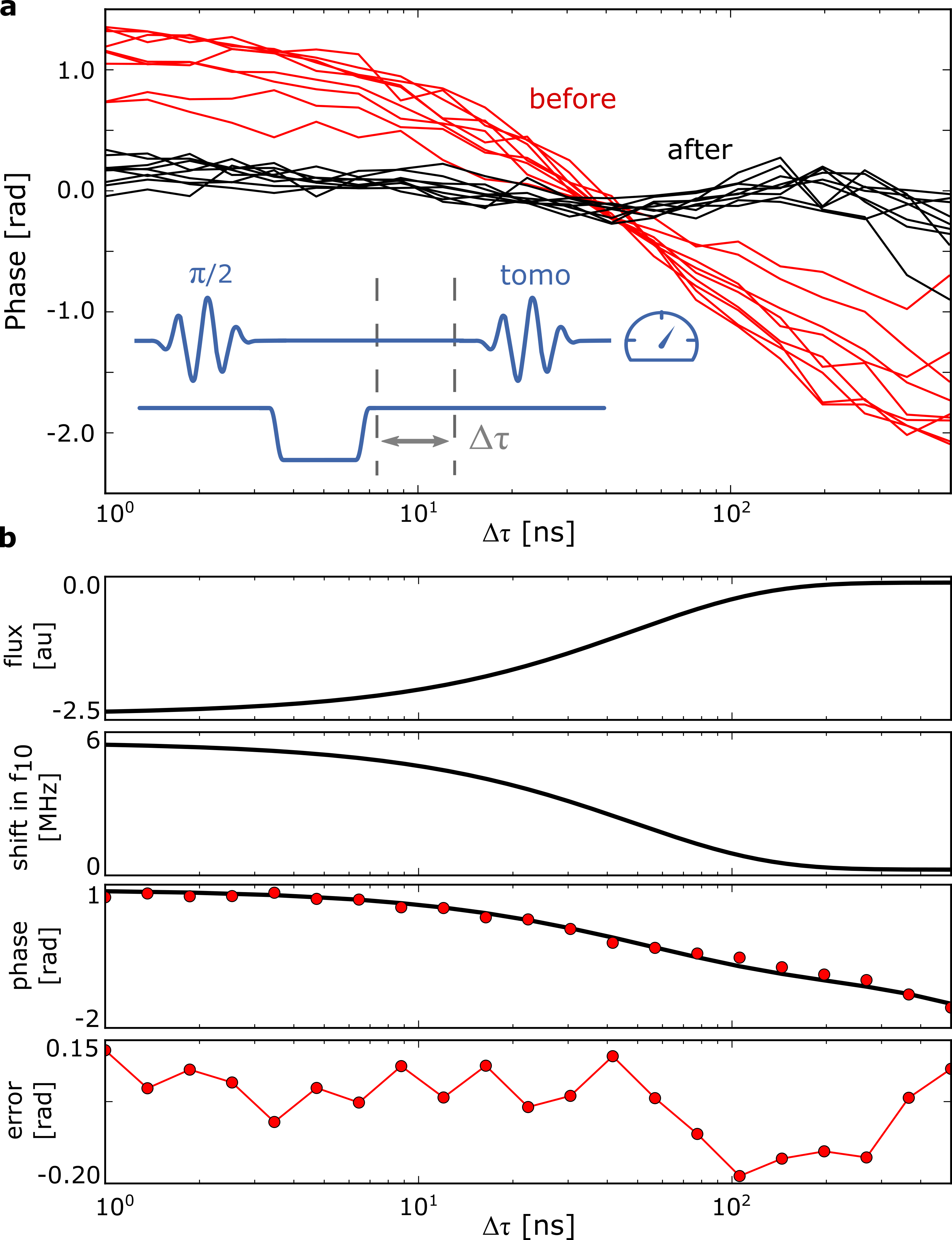}
	\caption 
	{
		\textbf{Pulse distortion.}
		\textbf{a} Qubit phase as a function of time after a pulse on the qubit flux-bias line.
		The ideal response is flat, however, due to imperfections in the lines, we observe the response shown in red.
		The data can be fit for the transfer function of the line and this can be used to flatten the response.
		The data after correction is shown in black.
		\textbf{b} Procedure for fitting the phase response data.
		Given a model for the transfer function, one can compute the expected flux versus time.
		Knowing the qubit frequency versus flux allows this to be converted to qubit frequency and then phase.
		This predicted phase can then be compared to the measured phase response and the difference provides a cost function for optimizing the parameters of the transfer function.
	}
	\label{fig:distortion}
\end{figure}

The pulse sequence used to infer the transfer function is shown inset in Fig.\,\ref{fig:distortion}a.  
First, the qubit is rotated to the equator.
Next, a square pulse is applied to the qubit flux-bias line. 
The phase of the qubit is then measured as a function of delay time after the pulse using state-tomography.
	Ideally, the phase should be independent of time.
However, we observe that the control pulse settles over time in a manner consistent with Eq.\,1.
The data is shown in red for all nine qubits.

The procedure used to fit the phase response is shown in panel \textbf{b}.
First, an initial guess for the parameters of the transfer function is used to predict the decay of the control pulse.
Next, the control flux is converted to qubit frequency using a separate calibration (data not shown).
The change in qubit-frequency can then be integrated to get the corresponding phase shift.
The difference between the predicted phase response and the measured phase response provides a cost function for finding the optimal parameters in the transfer function.
If one term is insufficient to flatten the response, a second amplitude and time-constant is added in order to further suppress phase accumulation.
The inferred transfer function can then be used to correct for the observed distortion by dividing the output signals ( in the frequency domain) by the transfer function.
The phase response after correction is shown in panel \textbf{a} (black).

\subsection{{Calibration 2:  Timing}}

Variations in the cabling and filtering can result in the pulses arriving at the target device at different times.
It is therefore important to measure these delays and offset the pulses in order to compensate.
Conceptually, this is done by choosing a single control line as a reference to which all other lines are synced.
In this experiment, the flux-bias line of the center qubit (Q5) is chosen to be the reference.
The rest of the lines are synchronized using the following steps.
1) The microwave pulses on Q5 are shifted so they align with the flux-bias pulses.
2) The flux-bias of the neighboring coupler (e.g. CP45) is shifted in order to sync with Q5's microwave line.
3) The flux-bias and microwave lines of Q4 are synchronized with one another by shifting the microwave pulse.
4)  Both the flux-bias and microwave lines of Q4 are shifted in order to sync up with CP45.
This procedure starts in the center and moves out towards both edges until the entire array is synced.

\begin{figure}[H]
	\centering
	\includegraphics{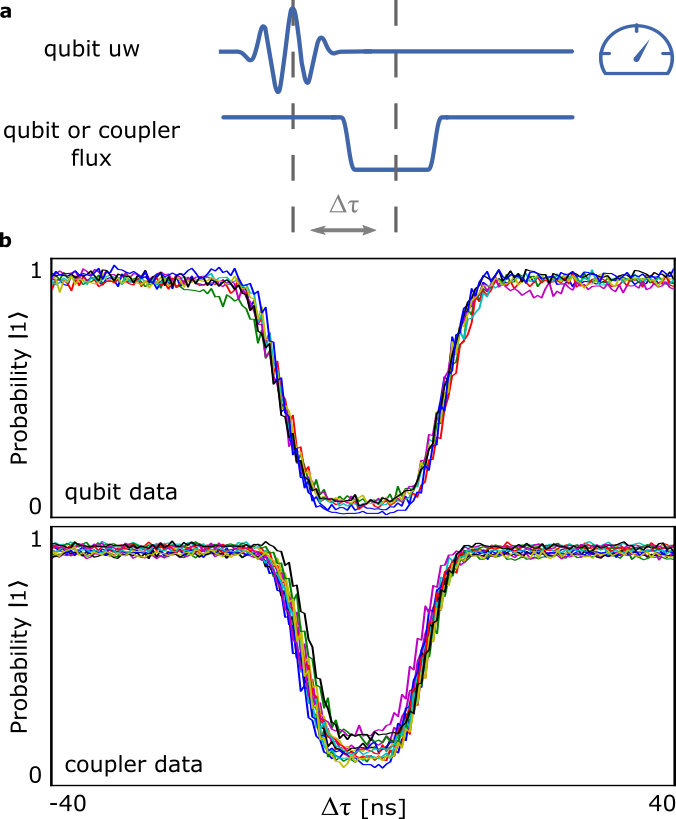}
	\caption 
	{
		\textbf{Timing delays.}
		\textbf{a} Pulse sequence used to synchronize timings for the 26 control lines.
		The qubit flux-bias lines and microwave lines are synchronized by varying the delay between a pi-pules and a detune pulse.
		When the relative timing is large, the pi-pulse excites the qubit; when the relative timing is small, the qubit is detuned from the microwave rotation and the pulse fails to excite the qubit.
		An identical experiment is used to synchronize the coupler flux-bias lines.
		This is done by using the frequency shift induced on the qubit by the coupler.
		\textbf{b} Experimental data for qubits and couplers.
		Each curve is fit to a sum of two error-functions in order to determine the timing offset between pulses.
	}
	\label{fig:timing}
\end{figure}

\begin{figure}[H]
	\centering
	\includegraphics{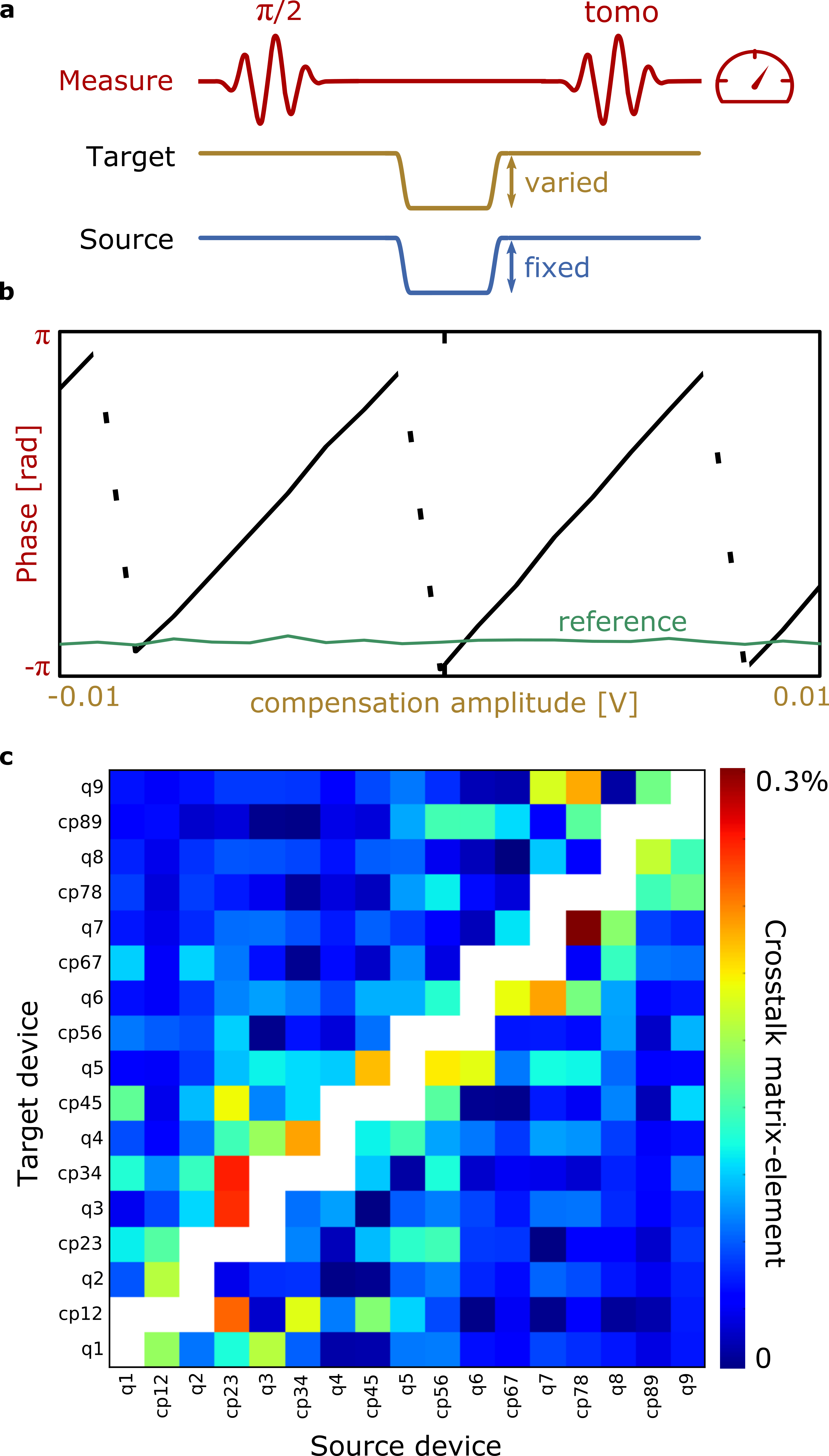}
	\caption 
	{
		\textbf{Flux-bias crosstalk.}
		\textbf{a, b} Pulse sequence and measurement used to calibrate cross-talk between flux-bias lines.
		This experiment is performed on all combinations of 'source' lines and 'target' qubits and couplers.
		For a fixed pulse height on the source line, we vary the height of a compensation pulse on the target line and measure the phase of a 'measure' qubit.
		The correct compensation amplitude corresponds to where the measured phase is equal to the phase without a pulse on the source line - this is the reference value (green) shown in the middle panel.
		The ratio of the compensation amplitude and the source pulse height is the crosstalk matrix-element.
		\textbf{c}  The cross-talk matrix.  Each pixel represent the crosstalk matrix-element from the source device to the target device.  Cross-talk from from qubits to neighboring couplers are not shown as they saturate the color-scale at 4\%.
	}
	\label{fig:crosstalk}
\end{figure}

The pulse sequence used to carry out this procedure is shown in Fig.\,\ref{fig:timing}a.
The relative timing between a microwave pi-pulse and a qubit (or coupler) flux-bias pulse is varied and the probability of the qubit being in the excited state is measured.
If the flux-bias pulse occurs before or after the pi-pulse, the qubit ends up in the excited state.
When the flux-bias pulse occurs during the pi-pulse, the qubit frequency shifts and the pi-pulse is off-resonance with the qubit and fails to completely excite the qubit.
Data for both qubit and coupler flux-bias pulses are shown in panel \textbf{b}.
The data is fit in order to determine the timing offset and the result is used to correct future pulses.

\subsection{{Calibration 3:  Crosstalk}}

Ideally, the current from any given control line will only reach a single device.
However, due to unwanted geometric coupling, a small fraction of the flux from any line will reach several devices.
The procedure used to measure and correct for this effect (linear flux crosstalk) is shown in Fig.\,\ref{fig:crosstalk}.

The pulse sequence and an example dataset are shown in panels \textbf{a} and \textbf{b}.
In order to measure the crosstalk, we apply a square pulse of fixed height on the 'source' line.
We then measure the phase of the 'measure' qubit versus the amplitude of the compensation pulse on the 'target' device.
The target device is either the measure qubit itself or a neighboring coupler.
The phase of the measure qubit in the absence of a pulse on the source line is measured as a reference.
The correct compensation amplitude corresponds to when the measured phase equals the reference phase.
The crosstalk matrix-element is given by the ratio of the amplitude of the compensation pulse and the source pulse.
This experiment is repeated for all combinations of sources and targets.

The crosstalk matrix-elements are shown in panel \textbf{c}.
Lines physically closest to one another have crosstalk around 0.1-0.3\%.
Crosstalk from qubits to neighboring couplers are not shown as they saturate the scale at 4\%.
The desired output fluxes are now multiplied by the crosstalk matrix in order to produce control fluxes with the ideal behavior.

\subsection{{Calibration 4:  Control model}}

\begin{figure}[t]
	\centering
	\includegraphics{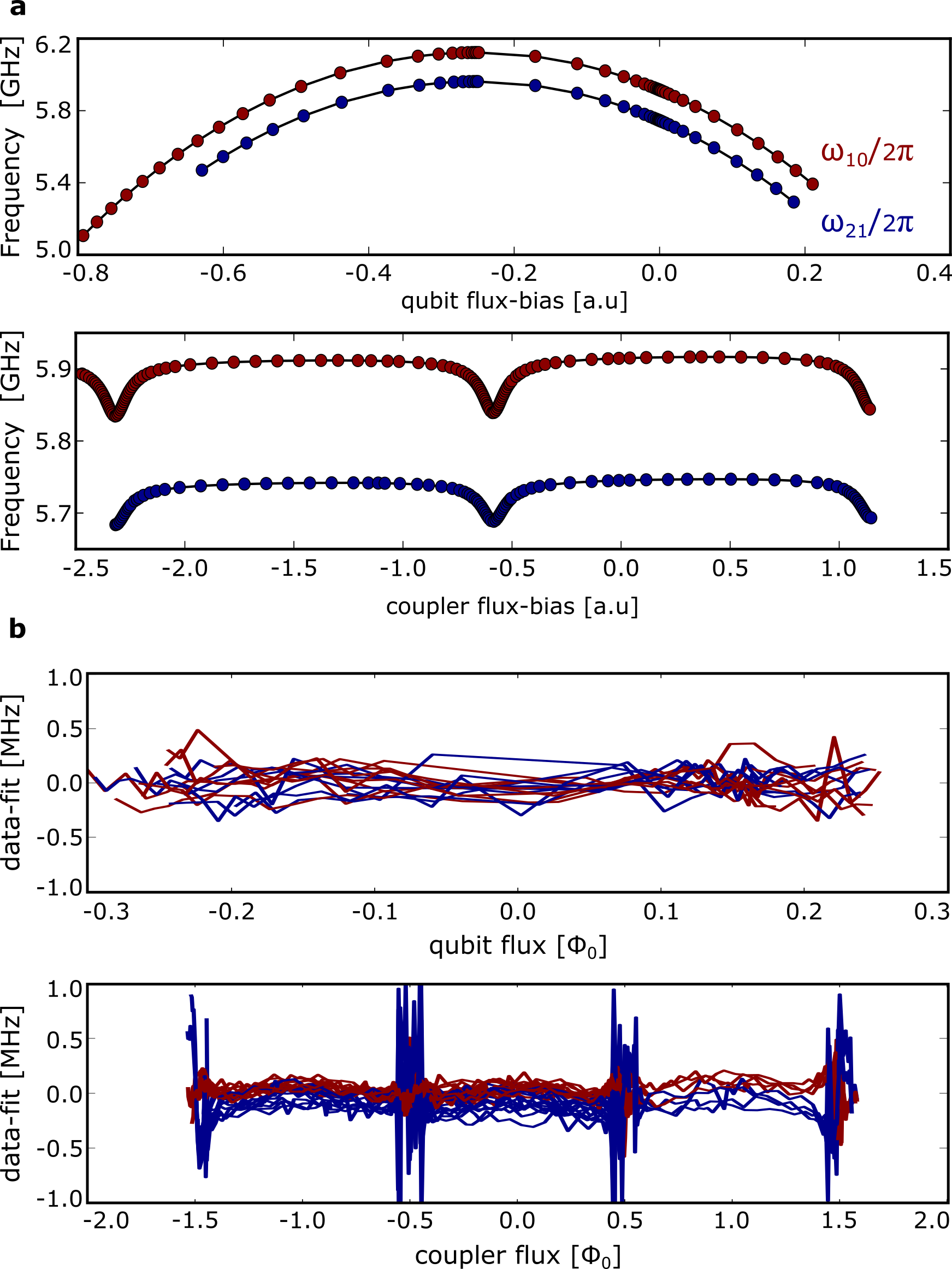}
	\caption 
	{
		\textbf{Building a physical control model.}
		\textbf{a, b} In order to convert the control pulses to Hamiltonian parameters, we construct a physical model of our qubits and couplers.
		The model consists of capacitors, inductors and Josephson junctions.
		In panel \textbf{a}, we measure the qubit transition frequencies $f_{10}$ (ground-state to first excited state) and $f_{21}$ (first excited to second excited state) as a function of qubit and coupler flux-biases.
		Fitting the datasets allows us to determine the values of model parameters.
		Fitting errors for all qubits and couplers are shown in panel \text{b} and are typically a few tenths of a MHz.
	}
	\label{fig:modeling}
\end{figure}

In the previous three sections, we've shown how to make sure that the control pulses are orthogonal, arrive at the same time, and arrive without distortion. 
This should provide us with sufficient knowledge of the pulses that arrive at the target device.
The next step in converting this control pulse to matrix-elements of a known Hamiltonian.

The qubits are well approximated by a Bose-Hubbard model with three parameters:  frequency, anharmonicity and coupling.
The frequency and anharmonicity have a nonlinear dependence on the flux-bias of the qubit and both neighboring couplers.
The coupling has a non-linear dependence on the flux-bias of the coupler, each qubit, and even the neighboring couplers (as they shift the qubit inductance).

Determining the values of these parameters as a function of the control knobs can be done in one of two ways:  1) measure every parameter versus every knob and spline the data or 2) measure cuts in the parameter space and use these cuts to construct a physical model of the device.
In this experiment, we have chosen the second approach for two reasons.
The first reason is that accurately measuring the individual parameters is very challenging.
Given two qubits, spectroscopic probes only provide two pieces of information (the two eigenvalues) from which one cannot simultaneously determine the frequencies and couplings separately (three parameters).
Time-domain approaches require accurately measuring small variations in the populations which are hard to resolve in experiments.
The second reason is that the final step in calibration is often optimization.
A 'measure everything and spline' approach has far too many degrees of freedom to optimize over.

A gmon qubit can be modeled as a capacitor, inductor and tunable junction, all in series.
The Hamiltonian for such a circuit is given by
\begin{equation}
\mathcal{H} = \frac{\hat{Q}^2}{2C} -\frac{I_0 \Phi_0}{2\pi}\cos{\hat{\phi}_j} 
+ \frac{1}{2}L_g\left(I_0\sin{\hat{\phi}_j}\right)^2
\end{equation}
where $\hat{Q}$ is the charge on the capacitor, C is the capacitance, $I_0$ is the tunable critical current of the junction, $\hat{\phi}_j$ is the phase drop across the junction, and Lg is the geometric inductance.
The phase drop across the junction $\hat{\phi}_j$ is related to the conjugate variable of charge (flux $\hat{\Phi}$) through the following relationship
\begin{equation}
\hat{\phi}_j = \hat{\phi} + 2\sum_{n}(-1)^n\frac{J_n(\beta n)}{n}\sin(n\hat{\phi})
\end{equation}
where $\hat{\phi} = \frac{2\pi}{\Phi_0}\hat{\Phi}$, $J_n$ is the Bessel function of the first kind, $\beta = L_g / L_j$, and $L_j = \frac{\Phi0}{2\pi} \frac{1}{I_0}$ is the effective inductance of the junction.
The Hamiltonian can be numerically diagonalized by expressing $\hat{Q}$ and $\hat{\Phi}$ in terms of harmonic oscillator raising and lowering operators.
We find that truncating these operators at 20 levels provides 1 Hz accuracy for a 5 GHz qubit with -200 MHz nonlinearity at $\beta = 0$.
It is important to note that keeping too many levels (past around 60) causes this numerical approach to break down as the harmonic oscillator states start to find low energy solutions in the neighboring minima of the cosine potential.

This model is used to fit measurements of the qubit energy spectrum for C, $L_j$ and $L_g$.
In order to make this process computationally efficient, exact diagonilization is used to estimate coefficients in a perturbation expansion; the perturbative model is then used for fitting.
This is done by expressing the Hamiltonian in terms of the harmonic oscillator frequency $\omega_0 = \frac{1}{\sqrt{C(Lg+Lj)}}$ and two small dimensionless parameters, $\beta = Lg/Lj$ and $\lambda = \frac{Z_0}{(R_k/\pi)}$ where $Z_0 = \sqrt{\frac{(Lg+Lj)}{C}}$ and $R_k = h/e^2$.
The Hamiltonian was diagonalized over a 100x100 grid for $0 \leq \lambda \leq 0.04$ and $0 \leq \beta \leq 0.25$ and the results were fit for coefficients of a 2-dimensional polynomial.
The following expressions for the two lowest transitions are accurate to within 100\,kHz at 5GHz
\begin{equation}\nonumber
\omega_{10}/\omega_0 = 1 + \sum A_{nm} \beta^n \lambda^{m+1}
\end{equation}
\begin{equation}\nonumber
\omega_{21}/\omega_0 = 1 + \sum B_{nm} \beta^n \lambda^{m+1}
\end{equation}
\begin{equation}\nonumber
A=
\begin{pmatrix}
-0.9989185  & -1.01547902 & -3.39493789 \\
2.92743183 & -1.15831188 & 0.0 \\
-4.93953913 & 8.17006907 & 0.0 \\
4.03181772 & 0.0 & 0.0
\end{pmatrix}
\end{equation}
\begin{equation}
B=
\begin{pmatrix}
-1.99707501  & -3.25782090 & -18.0220389 \\
5.81558214 & -1.77830584 & 0.0 \\
-9.55174679 & 22.6985133 & 0.0 \\
7.16401532 & 0.0 & 0.0
\end{pmatrix}
\end{equation}
Note that $A_{00} \simeq -1.0$ and $B_{00} \simeq -2.0$ as expected from first-order perturbation theory.
Similarly, $A_{10} \simeq 3.0$ (the coefficient of $\beta\lambda$) is also consistent with perturbation theory.
Increasing the domain or improving the accuracy of this expansion is as simple as fitting to a higher degree polynomial.

Each qubit is dispersively coupled to a readout resonator.
This resonator imparts a frequency dependent shift on the qubit's energy levels.
The frequency shift is modeled as
\begin{equation}
	\Delta \omega_{10} = \frac{1}{2} \left( \left| \delta \right|- \sqrt{4 g_r^2 + \delta^2} \right)
\end{equation}
where $\delta = \omega_{10} - \omega_{r}$, $\omega_{r}$ is the frequency of the resonator and $g_r$ is the qubit-resonator coupling.
The shift in $\omega_{20}$ depends on the anharmonicity of the qubit and is modeled as
\begin{equation}
	\Delta \omega_{20} = \frac{1}{2} \left( \left| \delta+\eta \right|- \sqrt{4 g_r^2 (1+\frac{\eta}{\omega_{10}}) + (\delta+\eta)^2} \right)
\end{equation}
where $\eta = \omega_{21} - \omega_{10}$ is the qubit anharmonicity - note that we have accounted for a nonlinear correction in the two-photon coupling strength.
The shift in $\omega_{21}$ is then given by $\Delta \omega_{21} = \Delta \omega_{20} - \Delta \omega_{10}$.
When fitting measurements of the qubit energy spectrum, these expressions are used to find $g_r$ given the known resonator frequency (measured independently).

Each qubit is coupled not only to a readout resonator but also to a coupler.
Treating the coupler as a tunable linear inductor, one finds that the geometric inductance of the qubit is modified according to the following expression
\begin{equation}
L_g \rightarrow L_g - \frac{M^2}{L_c} \frac{\beta_c}{1 + \beta_c}
\end{equation}
where M is the mutual inductance from the qubit to the coupler, $L_c$ is the geometric inductance of the coupler, $\beta_c = \beta_{C0} \cos \phi_c$ is the ratio of the coupler's geometric inductance to junction inductance, and $\phi_c$ is the phase drop across the coupler.
When converting from the applied control flux to junction phase, Eq.\,3 is used.
When fitting measurements of the qubit energy spectrum, these expressions are used to find $\beta_{C0}$ and $M_0 = M^2/L_c$.

The coupler also has a mode which dispersively pushes on the qubit energy levels (similar to the readout resonator) coming from the capacitance of the coupler junction.
We find that including this mode is necessary in order to accurately reproduce the measured qubit energy spectrum.
The frequency of this mode changes with the flux applied to the coupler according to 
$\omega_{c} = \omega_{C0} \sqrt{1+\beta_c}$ where $\omega_{C0}$ is the unbiased frequency of the coupler.
The coupling between the qubit and the coupler mode is given by $\frac{\sqrt{\omega_{10} \omega_c}}{2} \frac{M}{\sqrt{(Lg+Lj)Lc(1+\beta_c)}}$ which is the typical harmonic oscillator expression.
The dispersive shift is then taken into account in a manner similar to the readout resonator.
This effect introduces an additional fitting parameter $\omega_{C0}$.

All of these physical parameters are determined by fitting the two lowest transition energies of each qubit ($\omega_{10}$ and $\omega_{21}$) versus the qubit and coupler flux.
The advantage of this calibration strategy is that it requires only single qubit experiments and is likely scale to much larger systems.
Calibrating 9 qubits and 8 couplers takes around 24 hours to complete.
During each experiment, all other qubits are biased to their minimum frequency in order to effectively remove them from the system.
Ignoring the neighboring qubit could result in up to a 1 MHz error in the model at the maximum coupling (a 5\,GHz detuning and a 50\,MHz coupling gives a 0.5 MHz dispersive shift).
Example datasets are shown in Fig.\,\ref{fig:modeling}\textbf{a}.
The difference between the fit and the data is shown in panel \textbf{b} for all datasets.
Typical errors are on the order of 0.1\,MHz.
Below is a table of parameters inferred from the data.
$M_0$ left (right) refers to the mutual to the qubit on the left (right) of the coupler.

\begin{table}[H]
	\begin{center}
		\begin{tabular}{ | M{1.5cm} | M{1.5cm}| M{1.5cm} | M{2.0cm} | M{1.5cm} | } 
			\hline
			Qubit & C [Ff] & $L_j$[nH] & $L_g$ [nH] & $g_r/2\pi$ [MHz] \\
			\hline
			\hline
			Q1 & 86.2 & 6.46 & 0.96 & 112.0 \\ 
			Q2 & 85.9 & 6.26 & 0.98 & 106.0 \\ 
			Q3 & 87.7 & 6.25 & 0.86 & 135.0 \\ 
			Q4 & 86.5 & 6.47 & 0.92 & 128.0 \\ 
			Q5 & 83.9 & 6.26 & 1.06 & 106.0 \\ 
			Q6 & 85.6 & 6.31 & 0.98 & 114.0 \\ 
			Q7 & 85.9 & 6.45 & 0.95 & 113.0 \\ 
			Q8 & 86.4 & 6.33 & 0.94 & 117.0 \\ 
			Q9 & 87.2 & 6.41 & 0.86 & 126.0 \\ 
			Avg. & 86.1$\pm$1.0 & 6.35$\pm$0.08 & 0.95$\pm$0.06 & 117.4$\pm$9.5 \\
			Design & 85 & 7.0 & 0.9 & 110\\
			\hline 
			\hline
			Coupler & $M_0$ left [pH] & $M_0$ right [pH] & $\beta_{C0}$ & $\omega_{C0}/2\pi$ [GHz] \\
			\hline
			\hline
			CP12 & 43.6 & 40.2 & 0.664 & 14.6 \\ 
			CP23 & 42.6 & 41.1 & 0.660 & 14.6 \\ 
			CP34 & 42.3 & 41.6 & 0.665 & 14.8 \\ 
			CP45 & 43.2 & 41.5 & 0.661 & 14.7 \\ 
			CP56 & 46.2 & 40.9 & 0.657 & 15.0 \\ 
			CP67 & 44.0 & 41.8 & 0.664 & 14.8 \\ 
			CP78 & 43.3 & 39.5 & 0.671 & 14.5 \\ 
			CP89 & 43.2 & 37.9 & 0.663 & 14.8 \\ 
			Avg. & 43.6$\pm$1.1 & 40.6$\pm$1.2 & 0.663$\pm$0.004 & 14.7$\pm$0.15 \\
			Design & 50.0 & 50.0 & 0.7 & > 15 \\
			\hline
		\end{tabular}
	\end{center}
	\caption{Model parameters for all qubits (upper table) and couplers (lower table).
			The second-to-last row in each table is the average over devices $\pm$ 1 standard-deviation.  The last row is the design values.}
	\label{tab:modeling}
\end{table}

In addition to these, every control line has two more fitting parameters - a conversion from the full-scale DAC amplitude to $\Phi_0$ (2.04$\pm$0.03 for qubits and 1.82$\pm$0.04 for couplers) and a static flux offset.
The data is also fit for the crosstalk from couplers to qubits as this matrix-element is difficult to infer using any other technique.
Additional fitting parameters are included, when necessary, to account for the qubit being brought into resonance with two-level defects.
Lastly, we find it necessary to include a factor multiplying the two-photon interaction between the qubit and the coupler with inferred values of 0.959$\pm$0.003.
This 4\% reduction in the 2-photon coupling is likely the result of anharmonic corrections to the coupler.
Ideally, in future experiments, this parameter will be replaced by the coupler impedance and a nonlinear model of the coupler.

\section{Appendix}

\subsection{Complexity of the time evolution in the driven Hubbard model}

For a number of families of quantum circuits, such as linear optics (boson sampling)~\cite{AaronsonArkhipov2011}, commuting circuits (IQP)~\cite{bremner2015average,bremner16} and random circuits~\cite{BoixoCircuits,AaronsonChen2016}, it has been argued that sampling the distribution of the output of the circuit presents a hard computational task for a classical computer, see also Ref.~\cite{Lund2017} for a brief review. Also it has been suggested that these hardness arguments could be extended to a wider class of quantum circuits~\cite{Lund2017}. %It is expected that sampling from a distribution of outputs of unitary evolution with a sufficiently general ensemble of random evolution operators presents an exponentially hard problem~\cite{AaronsonArkhipov2011,AaronsonChen2016} see also Ref.~\cite{Lund2017} for a brief review. 
In this section we provide intuitive arguments in support of the computational complexity of the sampling of the output of the driven continuous evolution protocol implemented on the gmon circuit.  We show that the observed probability amplitudes $p_{n}$ of bitstrings $n=\{0,1\}^{\otimes N}$ can be mapped onto a classical partition function with complex temperature which realizes an analog of the sign problem of the Quantum Monte Carlo algorithm. %Calculating $p_{z}$ is a $\texttt{\#P}$ class problem. 
Computationally hard instances are likely generated in the quantum chaotic regime. To identify this regime we analyze characteristics of quantum chaos: the wave function statistics, rapid loss of memory of the initial state and entanglement dynamics. We restrict the discussion to the Bose-Hubbard approximation. The full gmon circuit model contains additional non-integrable terms in the Hamiltonian which do not change the qualitative picture of the chaotic dynamics. Further work is needed to formally establish the complexity class of the task of computing $p_{n}$ and to connect the sampling task in the gmon circuit to the collapse of the Polynomial Hierarchy, as was argued in the case of random circuits, see Ref.~\cite{BoixoCircuits} and references therein.

\subsubsection{Sampling amplitude as a classical partition function}

%(i) Heisenberg model mapping onto the classical partition function. 

%(ii) q-dits mapping onto classical partition function.

We separate the diagonal and even/odd terms in the Bose-Hubbard Hamiltonian,
\begin{align}
\mathcal{H}&=\hat{H}_d+\hat{V}_e+\hat{V}_o, \\
\hat{H}_d&=\sum _{i=1}^N\left( \delta _i \hat{a}_i^+\hat{a}_i +  \frac{\eta _i}{2}\hat{a}_i^+\hat{a}_i\left(\hat{a}_i^+\hat{a}_i-1\right)\right), \label{Ham}\\ \nonumber
\hat{V}_e&=\sum _{i\, \text{even}}g_{i,i+1}(t)\left(\hat{a}_i^+\hat{a}_{i+1}+h.c.\right), \\ \hat{V}_o&=\sum _{i \, \text{odd}}g_{i,i+1}(t)\left(\hat{a}_i^+\hat{a}_{i+1}+h.c.\right),
\end{align}
where $g_{i,i+1}(t)$ is the time-dependent drive as implemented in the experiment, see main text for details. For an evolution operator $\hat{U}(T)$ we introduce a grid of discrete time points $ t=0,...,2M$, related to the physical time $\tau$ by $\tau=\Delta t, \; \Delta\equiv \tfrac{T}{2M}$. We can write a Trotter decomposition in the basis of boson occupation numbers $|\vec{n}\left(t\right)\rangle \equiv \otimes_{i=1}^{N} |n_{i,t}\rangle$, where $n_{i,t}$ is the number of bosons at site $i$ at time $t$,
\begin{multline}
Z
\equiv 
\langle \vec{n}'|\hat{U}(T)|\vec{n}\rangle 
=\sum _{\{\vec{n}\left(t\right)\}}e^{-i \frac{1}{2}\Delta \sum _{t =1}^{2M}H_d\left(n\left(t\right)\right)} \nonumber \\
\times \prod_{t=0,2,4...}
\langle
\vec{n}\left(t\right)|e^{-i\Delta \hat{V}_e\left(t\right)}|\vec{n}\left(t
+1\right)\rangle \langle \vec{n}\left(t +1\right)| \\ e^{-i\Delta \hat{V}_o\left(t+1\right)}|\vec{n}\left(t +2\right)\rangle, 
\end{multline}
where the sum runs over all possible realizations of the set $\{\vec{n}\left(t\right)\}$.
All even (and odd) bond operators commute among themselves resulting in the product,
$
\langle e^{-i\Delta \hat{V}_{e}\left( t\right)} \rangle
= 
\prod_{i \, \text{even}} \Lambda_{(i,i+1); (t,t+1)},
$
where each term $\Lambda_{(i,i+1); (t,t+1)}$ is
\begin{align}
\langle n_{i, t}n_{i+1, t}|  e^{-i \Delta g_{i,i+1}\left(t\right) (\hat{a}_i^{\dagger }\hat{a}_{i+1} + \hat{a}_{i+1}^\dagger \hat{a}_i)}|n_{i, t+1} n_{i+1, t +1}\rangle\;.\nonumber
\end{align}
% which can be calculated explicitly recalling the definition of ladder operators, $
% \hat{a}_i^{\dagger }=\sum _{n_i=0} \sqrt{n_i+1}|n_i+1\rangle \langle n_i|,\;
% \hat{a}_i=\sum _{n_i=1}^{\infty } \sqrt{n_i}|n_i-1\rangle \langle n_i|$,
This can be calculated explicitly giving % $\Lambda_{(i,i+1); (t,t+1)}

\vspace{.7em}
\begin{tabular}{|L|L|}
	\hline 
	\mathbf{\Lambda_{(i,i+1); (t,t+1)}} & \text{ \bf{Condition}} \\
	\hline
	1 & \begin{tabular}{@{}c@{}} $n_{i,t}=n_{i,t+1} $\\$ n_{i+1,t}=n_{i+1,t+1}$ \end{tabular}\\
	\hline
	-i \Delta g_{i,i+1}\left(t\right) \sqrt{n_{i,t} (n_{i+1,t}+1)}  & \begin{tabular}{@{}c@{}} $ n_{i,t}=n_{i,t+1}+1$ \\ $ n_{i+1,t}=n_{i+1,t+1}-1 $ \end{tabular}\\
	\hline
	-i \Delta g_{i,i+1}\left(t\right) \sqrt{(n_{i,t}+1) n_{i+1,t}}  &\begin{tabular}{@{}c@{}} $  n_{i,t}=n_{i,t+1}-1$ \\ $ n_{i+1,t}=n_{i+1,t+1}+1 $ \end{tabular} \\
	\hline
	0 & \text{other} \\
	\hline
\end{tabular}
\vspace{.7em}

This operator reflects a swap of one particle between the neighboring sites $n_{i,t}$ and $n_{i+1,t}$ and ensures that particle non-conserving trajectories do not contribute to the partition function. The partition function is a sum of complex amplitudes,
\begin{gather}
Z=\sum _{\{\vec{n}\left(t\right)\}} e^{i \phi_{\{\vec{n}\left(t\right)\}}} w_{\{\vec{n}\left(t\right)\}}. \label{ZphaseXweight}
\end{gather}
Each trajectory $\{\vec{n}\left(t\right)\}$ is associated with a phase factor $\phi_{\{\vec{n}\left(t\right)\}}$, and a real non-negative weight $w_{\{\vec{n}\left(t\right)\}}\geq0$,
\begin{align}
w_{\{\vec{n}\left(t\right)\}}&=%\prod _{(i,t)\in \mathcal{M}_{dw}}\left| \cos \left(g_{i,i+1}\left(t\right)\Delta\right)\right| 
\nonumber\\ 
& \prod _{\text{type}\; I}%{n_{i,t}=n_{i,t+1}+1,n_{i+1,t}=n_{i+1,t+1}-1,}%{(i,t)\in \mathcal{N}_{dw}} 
\left| \sqrt{n_{i,t} (n_{i+1,t}+1)} g_{i,i+1}\left(t\right)\Delta\right| \nonumber\\ &
\prod _{\text{type}\; II}%{(i,t)\in \mathcal{N}_{dw}} 
\left| \sqrt{(n_{i,t}+1) n_{i+1,t}} g_{i,i+1}\left(t\right)\Delta\right|,\nonumber
\end{align}
where the products are over all 4-site plaquettes  of type-$I$ and type-$II$ for which the equalities $ n_{i,t}=n_{i,t+1}+1, n_{i+1,t}=n_{i+1,t+1}-1$ and $ n_{i,t}=n_{i,t+1}-1, n_{i+1,t}=n_{i+1,t+1}+1$, respectively, hold for a given 2D trajectory $\{\vec{n}\left(t\right)\}$. According to the expressions for $\Lambda_{(i,i+1); (t,t+1)}$ the weight $ w_{\{\vec{n}\left(t\right)\}}$ vanishes for any trajectory that does not conserve the number of bosons.

The phase factor, from the swaps and from the diagonal part of the Hamiltonian, is
\begin{multline}
\phi_{\{\vec{n}\left(t\right)\}} =
-\frac{1}{2} \Delta \sum _{t =1}^{2M} \sum _{i =1}^{N} \left( \delta_i n_{i,t}+\frac{\eta}{2}n_{i,t}(n_{i,t}-1)\right)
\\- \frac{\pi}{2}\sum_{(i,t)} J_{(i,i+1);  (t, t+1)} \left(n_{i,t} - n_{i,t+1} \right)^2 \\\left(n_{i+1,t} - n_{i+1,t+1}\right)^2, \label{Zphase}
\end{multline}
where $J_{(i,i+1);  (t, t+1)}$ $= $ $\mathrm{sign} \left(g_{i,i+1}(t)\right)$ when $ n_{i,t}$ $=$ $ n_{i,t+1}\pm1$ and $ n_{i+1,t}$ $=$ $ n_{i+1,t+1}\mp1 $, but is equal to $0$ otherwise. 

The resulting partition function $Z$ in Eqs.~(\ref{ZphaseXweight}-\ref{Zphase}) describes a discrete classical model, a generalization of the Potts model~\cite{RevModPhys.54.235} on a square lattice with 4-site interactions and complex parameters with non-zero real and imaginary parts.

\begin{figure}%[H]
	\centering\includegraphics[width=.8\columnwidth]{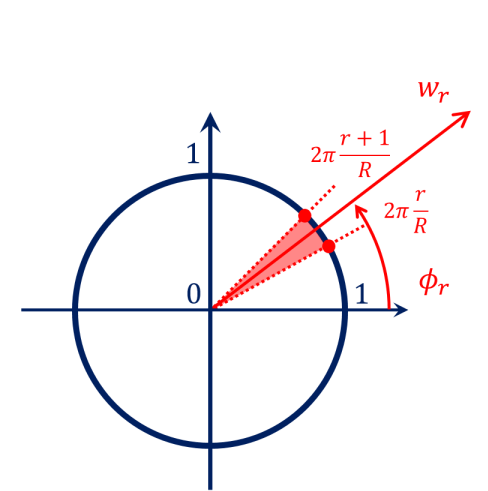}
	\caption{Binning of the phase factors in the partition function Eq.~(\ref{complexZ}).}\label{fig:sign_problem}
\end{figure}

Assuming a classical algorithm with overall polynomial resources, the accuracy with which each phase factor of a trajectory can be determined is polynomial in $N$. This means that the partition function takes the form, 
\begin{gather}
Z=\sum_{r=0}^{R}  w_r e^{2\pi i \frac{r}{R}},\label{complexZ}
\end{gather}
where $r$ is integer, and $R\sim \mathcal{O}(N^\alpha)$ for some power $\alpha$. Each weight $ w_r$ is therefore a bin containing all the trajectories corresponding to the given phase within the discretization accuracy, see Fig.~\ref{fig:sign_problem}, i.e. $w_r$ is a sum $  w_r=\sum_{i=0}^{\kappa_r} w_{r}^{(i)}$ over $\kappa_r$ trajectories satisfying the condition,
\begin{gather}
2\pi  \frac{r}{R} <\phi_{\{\vec{n}\left(t\right)\}}<2\pi \frac{\left(r+1\right)}{R}, \label{rphase}
\end{gather} 
each trajectory weighted with $ w_{r}^{(i)}$. The typical number of trajectories satisfying Eq.~(\ref{rphase}) is exponential in the number of time steps multiplied by the number of sites $M\times N$. Notice however that in the limit $M\rightarrow\infty$ the weight of each trajectory is suppressed because $w_{\{\vec{n}\left(t\right)\}} \sim \exp\left(-\nu \log M\right)$, where $\nu$ is the total number of swaps of both type-I and type-II in the trajectory. For a given realization of a trajectory  $\{\vec{n}\left(t\right)\}$ containing $\nu$ swaps in a specific order there are $ {M}\choose{\nu}$ ways to arrange the swaps among the $M$ points of discrete time. This entropic factor compensates the suppression factor in the weight of the trajectory. As a result, trajectories with $ \nu_{gT}  \sim  N \int_0^{T}dt g_{i,i+1}(t) \gg1$, which does not scale with $M$, dominate the weight $w_r$. Therefore $w_{r,\nu_{gT}}\sim \exp\left( \textrm{const} \times \nu_{gT}\right) $, which ensures that $w_r \gg 1$ scales exponentially with the number of qubits $N$. The partition function is therefore given by the sum of a polynomial number of phase factors each multiplied by an exponentially large weight.

In the chaotic regime the typical value of the amplitude $\left|\langle \vec{n}'|\hat{U}(T)\vec{n}\rangle  \right|$ is exponentially small in the number of qubits $N$, and therefore the value of $\left|Z\right|$ results from the cancellation of exponentially large weights. In this regime the task of calculating the sum in Eq.~(\ref{complexZ}) on a classical computer %is in the class $\texttt{\#P}$, which 
is analogous to estimating a partition function with a sign problem using the Quantum Monte Carlo algorithm. In Ref.~\cite{BoixoCircuits} it was argued that a classical statistical algorithm to calculate the sum of type Eq.~(\ref{complexZ}) in polynomial time necessarily requires exponential accuracy, suggesting that approximating Eq.~(\ref{complexZ}) presents a hard computational problem.

\subsubsection{Comparison to the cases of hard core and free bosons}\label{sec:free}

It is instructive to consider the case where the Hilbert space is truncated to only the ground and first excited level $n_{i}=\{0,  1\}$, which holds in the limit $\eta \to \infty$. This corresponds to the spin-$1/2$ XX model. In this case the Hubbard interaction does not contribute and therefore the evolution of the Bose-Hubbard model at half-filling can be described in terms of the dynamics of $N/2$ free particles. Moreover, the model allows a one-to-one mapping onto free fermions via the Jordan-Wigner transformation~\cite{NagaosaBook}. 
The initial state can be written in terms of fermion creation/annihilation operators $c_i^\dagger$, $c_i$ at lattice site $i$,
\begin{gather}
|\vec{n}\rangle =\prod_{\alpha=1}^{N/2} c^{\dagger}_{i_\alpha}  |0\rangle, \;\;\;  1\leq i_1 < i_2<...<i_{N/2}\leq N,
\end{gather}
where strict ordering of indexes means Jordan-Wigner strings can be dropped. Rewriting the time-dependent Hamiltonian Eq.~(\ref{Ham}) in terms of fermionic operators and the time dependent $N\times N$ matrix $\hat{h}$, $\mathcal{H}=\sum_{i,j} \hat h_{i,j}(t) c^{\dagger}_{i} c_{j} $, we can find the time dependence of the Heisenberg picture fermionic operators explicitly,
\begin{align}
\tilde{c}_i(T)&=\hat U^\dagger(T)  c_j(0)\hat U(T) =\sum_{j=1}^N \left[V_{i, j}(T) \right] c_j(0), \label{eq:heisenberg_fermion}\\
\left[ V_{i, j}\left(T\right) \right]   &= \left[ \mathcal{T}\exp\left(- i  \int_0^{T} dt\, \hat{h} (t) \right) \right]_{ij},
\end{align}
where $V_{i, j}$ is an $N\times N$ matrix. The anti-commutation properties of free fermions result in the fully anti-symmetric form of the probability amplitude, 
\begin{multline}
\langle \vec{n}'|\hat{U}(T)|\vec{n}\rangle=\langle 0|\prod_{\beta=1}^{N/2} \tilde c_{j_\beta}(T) \prod_{\alpha=1}^{N/2} c^{\dagger}_{i_\alpha} |0\rangle  \\= \sum_{\mathcal{P}(i_{\alpha})} (-1)^\mathcal{P}V_{j_1, i_1}\left(T\right) V_{{j_2},{i_2}}\left(T\right) ... V_{{j_{N/2}}, {i_{N/2}}}\left(T\right),  \label{Zxx}%\\ 
\end{multline}
where $\mathcal{P}(i_{\alpha})$ stands for permutations of the site indexes $i_{\alpha}$ and the factor $(-1)^\mathcal{P}$ takes into account the sign change for odd number of index permutations. The sum of $\left(\tfrac{N}{2}\right)!$ terms in Eq.~(\ref{Zxx}) reduces to a determinant of an $\tfrac{N}{2}\times\tfrac{N}{2}$ matrix and can be calculated efficiently. This is a very special realization of the partition function Eqs.~(\ref{ZphaseXweight}-\ref{Zphase}). 

In the opposite limit of $\eta\rightarrow 0$ and unlimited Hilbert space $n_i={0, 1, ..., \tfrac{N}{2}}$, a similar description holds in terms of free bosons. In particular, the Heisenberg equation of motion is completely analogous to Eq.~\eqref{eq:heisenberg_fermion} but using bosonic operators $a_j(T)$ instead of fermionic operators. Crucially, this implies that the amplitude $\langle \vec{n}'|\hat{U}(T)|\vec{n}\rangle$ does not have the form of an anti-symmetric sum, but instead it can be reduced to a permanent~\cite{scheel2004permanents}, which is known to present a hard computational problem~\cite{valiant1979complexity,lipton1989new,AaronsonArkhipov2011}. Note also that because Eq.~\eqref{eq:heisenberg_fermion} can swap operator indexes as part of the evolution, the restriction to 1D is not consequential for $T \in O(N)$. 

In the intermediate regime of $\eta/g \in O(1)$, including the second on-site excited level $n_{i}=\{0, 1, 2\}$ violates the strictly anti-symmetric nature of the sum in Eq.~(\ref{Zxx}). In the case of weak $\eta$, an intuitive picture emerges: virtual transitions through the second excited state introduce effective interactions in the fermionic representation. For random parameters, we expect the interactions to result in a chaotic evolution. 

\subsection{Estimates for the cost of simulating the Bose-Hubbard model.}

\subsubsection{Direct simulation with truncated bosons}\label{sec:app_direct_truncation}

A first attempt to estimate the cost of a direct simulation of the Bose-Hubbard model in the parameter range of interest is to truncate the Fock space of each gmon. 
We denote the different truncated spaces by the parameter $m$, with $m=1$ corresponding to all the gmons truncated to the first excitation level $\ket 1$ (the qubit subspace), $m=2$ corresponding to  all the gmons truncated to the second excitation level $\ket 2$, and so for. 

We integrate the time-dependent Schr\"odinger equation using fourth order Runge-Kutta, which is a standard numerical integration method~\footnote{There are other numerical integration algorithm, such as the Lanczos method. Nevertheless, given that we are interested in a time-evolution with a time-dependent Hamiltonian, fourth order Runge-Kutta seems well suited for this task.}.  We use the cross entropy difference after projecting the state to the qubit subspace, as explained in the main text, as an approximation of fidelity, to quantify the fitness of the numerical integration at different levels of truncation. More explicitly, we performed numerical integration with up to $N=18$ gmons and compute the cross entropy difference at a given truncation level $m$ and the fiducial integration with the highest truncation level, which we choose to be $m=4$.

\begin{figure}%[H]
	\centering\includegraphics[width=\columnwidth]{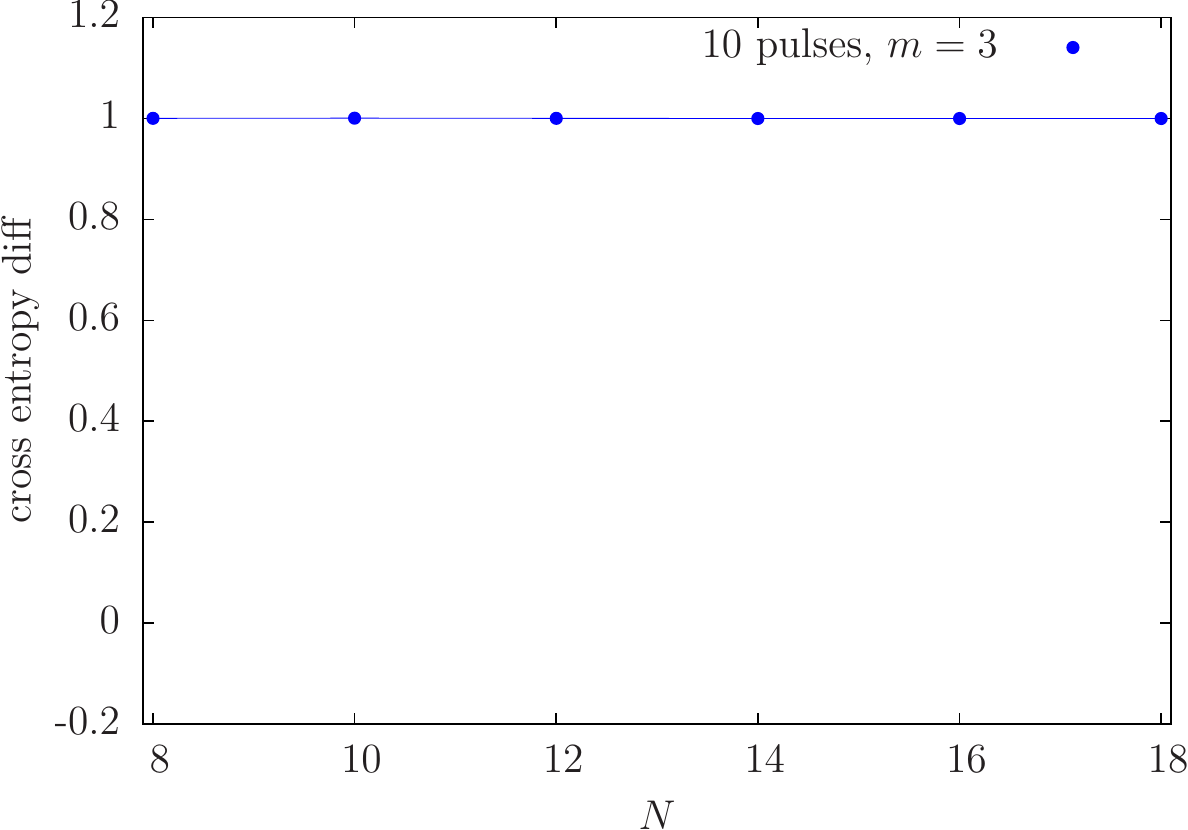}
	\caption{We can simulate with four level systems.}\label{fig:m3ce}
\end{figure}

\begin{figure}%[H]
	\centering\includegraphics[width=\columnwidth]{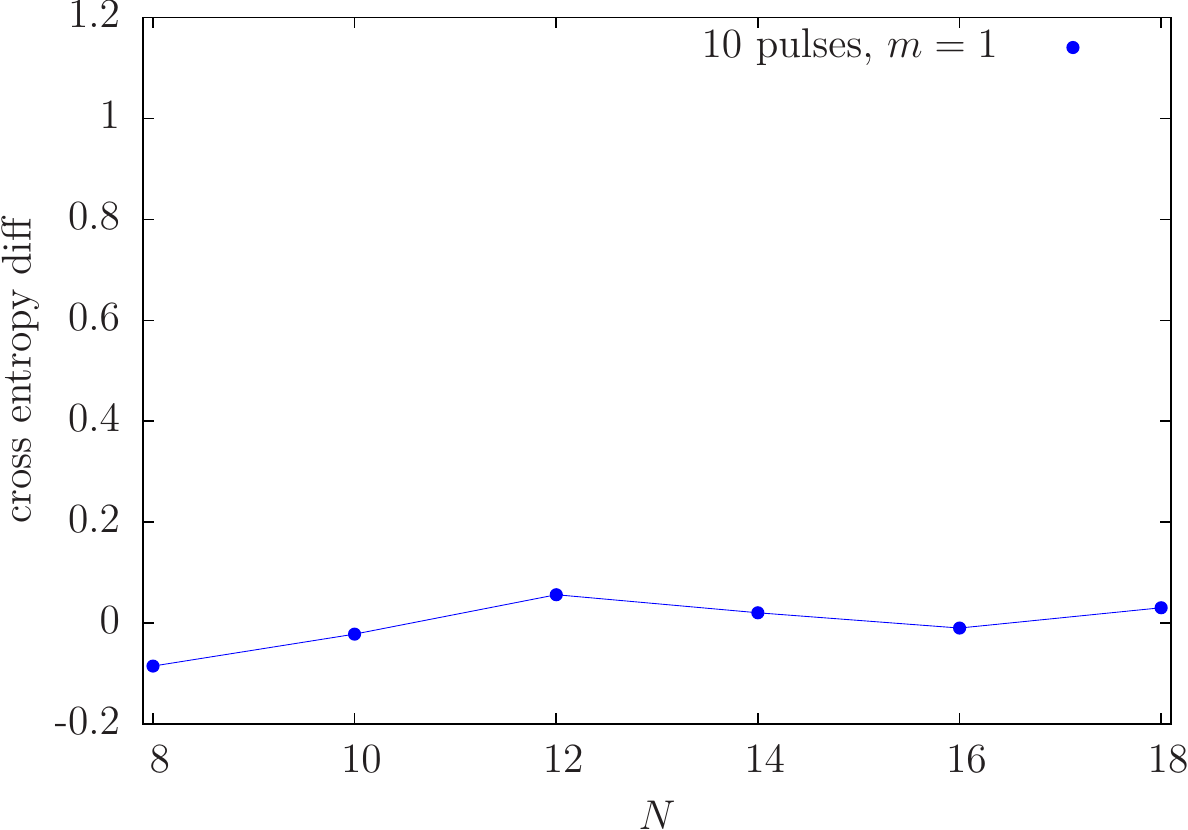}
	\caption{We can not simulate the system with plain qubits.}\label{fig:m1c3}
\end{figure}

\begin{figure}%[H]
	\centering\includegraphics[width=\columnwidth]{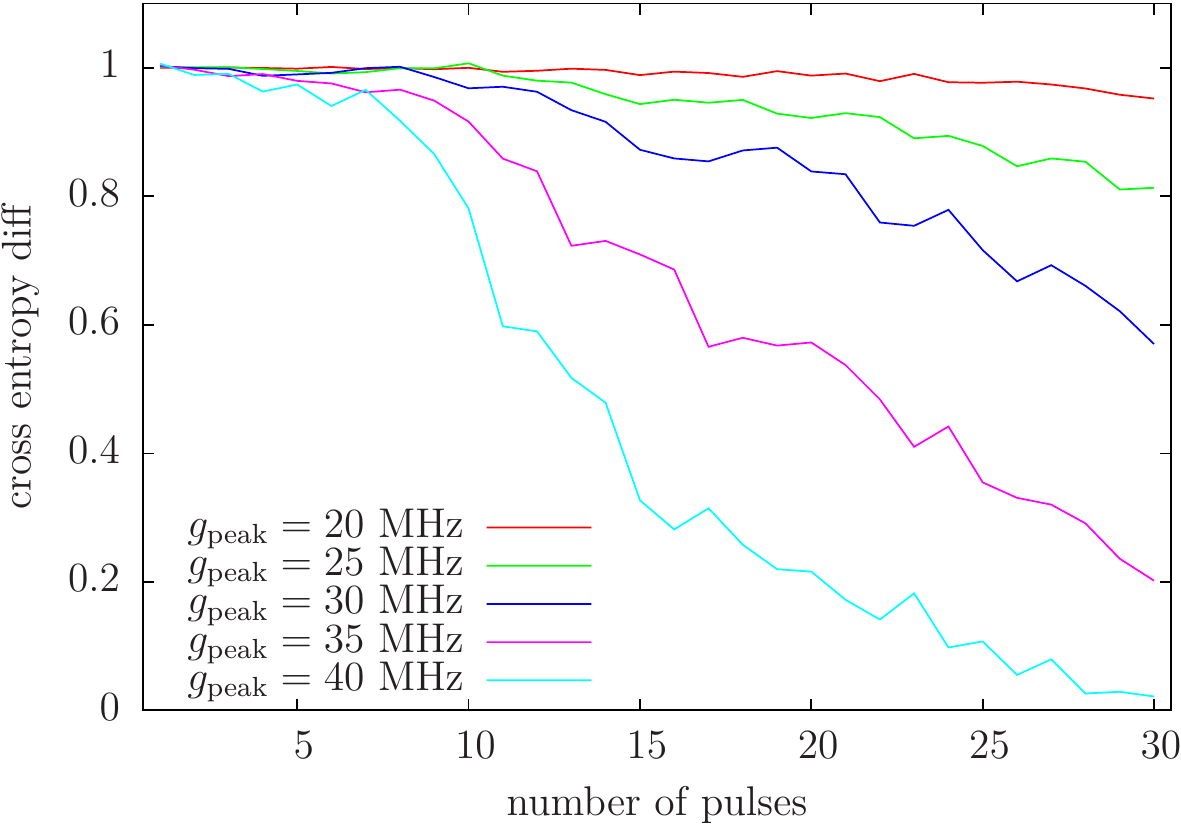}
	\caption{Cross entropy difference with three level systems for 16 gmons, $m=2$ and pulse duration $T_\text{pulse} = 20.5 \pm 4.5$ ns (the duration is chosen uniformly at random in the interval of 16 to 25 ns).}\label{fig:m2c3}
\end{figure}

Our first observation is that the cross entropy difference between $m=3$ and $m=4$ is exactly 1, see Fig.~\ref{fig:m3ce}. This implies that a direct numerical numerical integration with $m=3$ (or $m=4$) is sufficient to simulate the dynamics of the Bose-Hubbard model at the parameter regime of interest, because at this point including more energy levels does not translate in any difference in the resulting state. Our second observation is that a direct truncation to the $m=1$ subspace is not a valid approximation, resulting in a fidelity (cross entropy difference) $\sim 0$, see Fig.~\ref{fig:m1c3}. This is to be expected, because this approximation can be mapped to free fermions, which does not result in chaotic dynamics (see Secs.~\ref{sec:free} and~\ref{sec:chaotic_state}).
Figure~\ref{fig:m2c3} shows the cross entropy difference (fidelity) for 16 gmons as a function of the number of pulses for representative parameters with a truncation to the second excited state $\ket 2$ of each gmon, $m=2$. We see that the fidelity decreases as $g$ increases, because higher energy states become increasingly important (see also Fig.~\ref{fig:exact_spectrum}). Nevertheless, the decay of fidelity is likely not worse than the experimental fidelity. Therefore, this is a valid simulation method for the purposes of comparing the cost of a classical simulation against the experimental implementation.

The effective Hilbert space size of the  truncation to the second excited state $\ket 2$ of each gmon, $m=2$, is $3^m$, if we do not take into account boson number conservation. Taking into account boson number conservation, we calculate the resulting Hilbert space exactly, at half filling, for a number of gmons $N$ between $N=20$ and $N=40$, which is the regime of interest for a future comparison between experiment and classical algorithms. We fit the resulting curve, an obtain a good fitting for the effective Hilbert space dimension $D \sim 2.4^N / 0.14$ (see also Fig.~\ref{fig:mt_estimates}).

\subsubsection{Estimate of Runge-Kutta integration steps}

\begin{figure}%[H]
	\centering\includegraphics[width=\columnwidth]{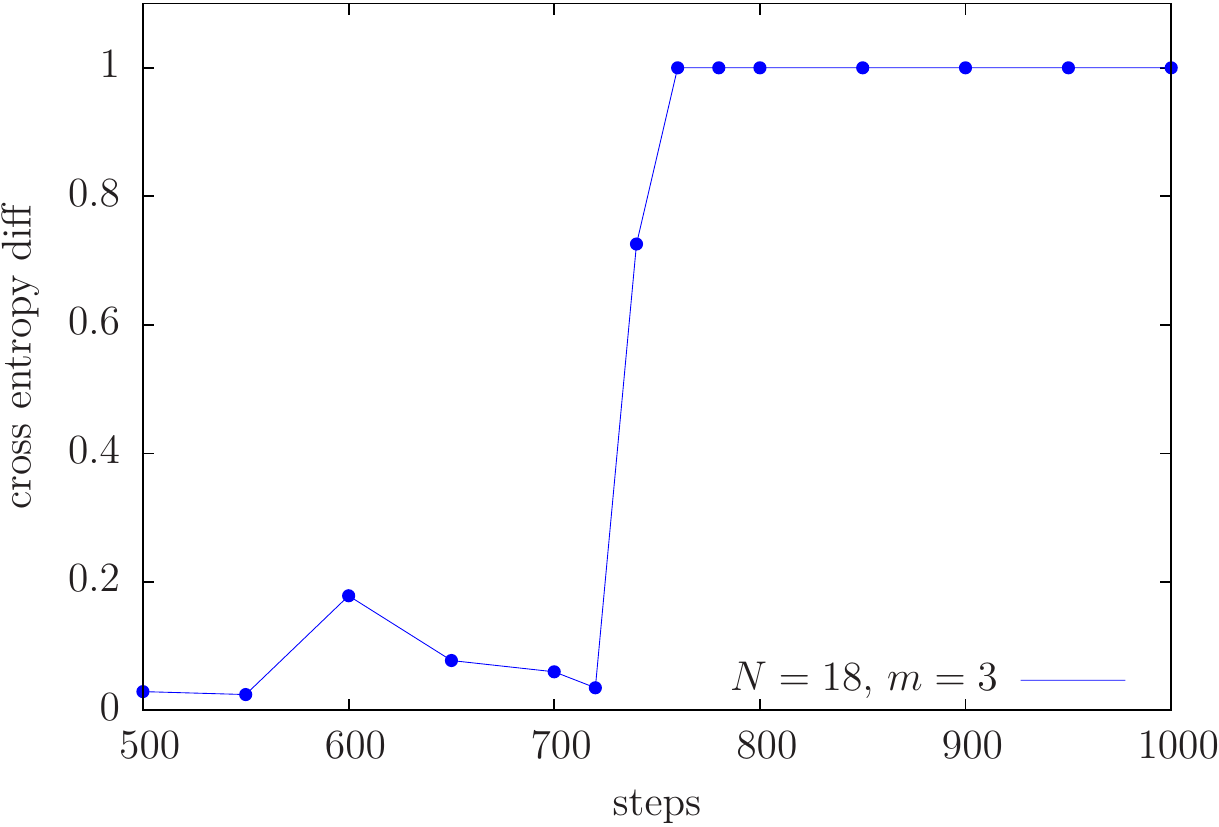}
	\caption{Cross entropy as a function of the number of Runge-Kutta steps for 18 gmons, 5 pulses and $T_\text{pulse} = 30 \pm 10$ ns.}\label{fig:ce_steps}
\end{figure}

\begin{figure}%[H]
	\centering\includegraphics[width=\columnwidth]{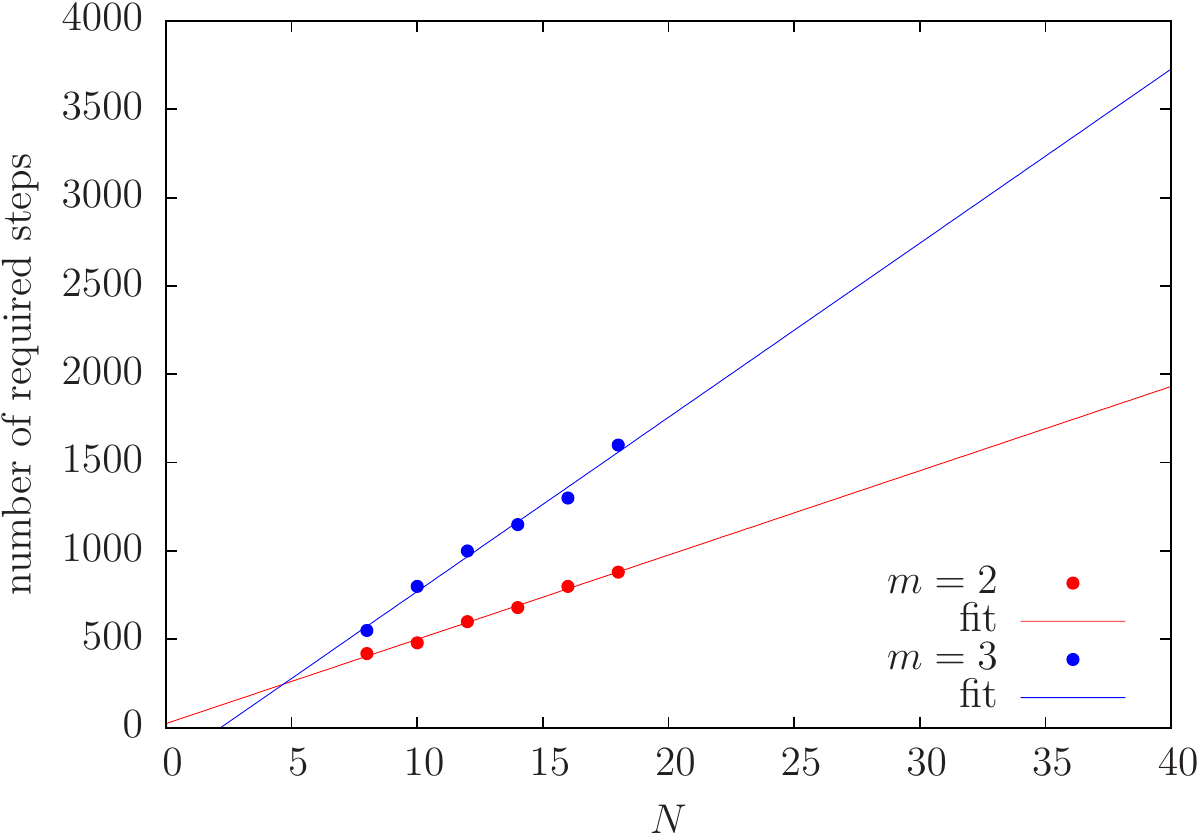}
	\caption{Required number of Runge-Kutta integration steps for 10 pulses and $T_\text{pulse} = 30 \pm 10$ ns.}\label{fig:rk_required_steps}
\end{figure}

We now provide an estimate for the number of fourth order Runge-Kutta integration steps required for a numerical simulation. At the end of the integration, we project the resulting state into the qubit subspace expanded by the states $\{\ket 0, \ket 1\}$ of each gmon (the first band, see Sec.~\ref{sec:app_bands}). It can be seen that we can tolerate large errors in the populations of states with higher excitation numbers (which are thrown away in the final measurement), while the populations in the qubit subspace are still accurate.

We observe a sensitive behavior of the numerical integration error in the qubit subspace for increasing number of integration steps. Figure~\ref{fig:ce_steps} shows the cross entropy for $18$ gmons and 5 pulses ($T_{\rm pulse} = 30 \pm 10$ ns) as a function of the number of integration steps. We see that if the number of steps is less than $\sim 700$ the numerical integration is completely inadequate. At the same time, we don't obtain any benefit using more than $\sim 800$ integration steps. 

This steep behavior is used to estimate the required number of integration steps for increasing number of gmons $N$, instead of using an adaptive stepsize Runge-Kutta scheme. Figure~\ref{fig:rk_required_steps} gives estimates for 10 pulses ($T_{\rm pulse} = 30 \pm 10$ ns) as a function of $N$. We show estimates of numerical integrations truncating at the second excited state $\ket 2$ ($m=2$), and the third excited state $\ket 3$ ($m=3$). %For the purposes of estimating the cost of a classical simulation, we would use an estimate of $1000$ to $2000$ Runge-Kutta integration steps for 10 pulses. 

\subsubsection{Band structure}\label{sec:app_bands}

\begin{figure}%[H]
	\centering\includegraphics[width=\columnwidth]{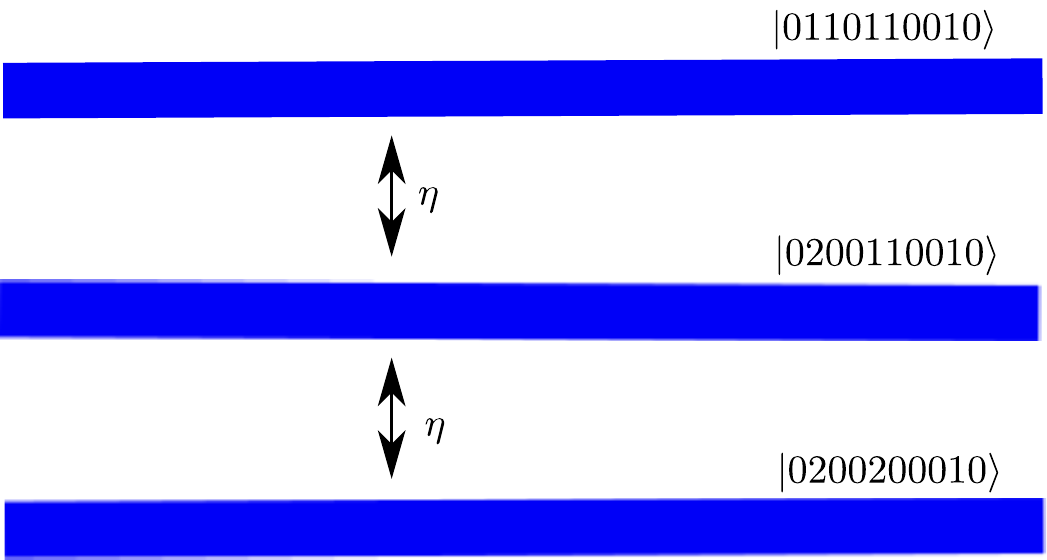}
	\caption{Symbolic representation of state bands, separated by the non-linearity $\eta$.}\label{fig:bands}
\end{figure}

In the limit $g \ll \eta$ of the swapping parameter $g$ much smaller than the non-linearity $\eta$, the spectrum is separated by bands characterized by the occupation number of the different gmons, see Fig.~\ref{fig:bands}. The first band is the qubit subspace, with states composed by superpositions of Fock states where each gmon is either in the ground state $\ket 0$ of the first excited state $\ket 1$. At half filling, and using the fact that the Bose-Hubbard model conserves the total number of bosons, the Hilbert space dimension of this band is
\begin{align}
{N \choose  {N/2}} \sim \frac {2^N} {\sqrt {\pi N /2}}\;.
\end{align}
The next band, separated by an energy $\eta$, contains states composed by superpositions of Fock states where exactly one gmon is in the second excited state or ``doublon'' $\ket 2$. At half filling, its Hilbert space dimension is given by the multinomial coefficient
\begin{align}
{N \choose N/2-1,N/2-2,1}\;.
\end{align}
The next band after that has states with exactly two gmons in the ``doublon'' state $\ket 2$, etc...

\begin{figure}%[H]
	\centering\includegraphics[width=\columnwidth]{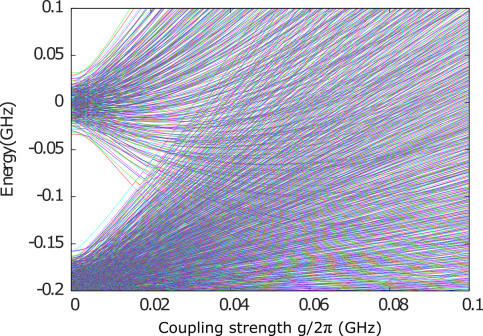}
	\caption{Exact diagonalization with $10$ gmons.}\label{fig:exact_spectrum}
\end{figure}

Our next observation is that the bands overlap for increasing $g$. Figure~\ref{fig:exact_spectrum} shows the exact spectrum for the Bose-Hubbard model and increasing $g$. We see that at $g \sim 20$ MHz the bands start to overlap. One implication for estimating the cost of classical simulations is that perturbation theory in the qubit subspace, such as the Schrieffer-Wolff transformation, is not expected to work. This method is an expansion in the perturbation parameter $g/\eta$, and assumes that the bands remain well separated. 

\subsubsection{Simulation with a fixed number of bands}\label{sec:app_fixed_bands}

In Sec.~\ref{sec:app_direct_truncation} we have seen that for a direct numerical simulation it suffices to truncate each gmon to the second excited state $\ket 2$ ($m=2$ truncation in our notation). Using the band structure explained in the previous section, we now outline how an approximate classical simulation can be carried out more efficiently. Given that the Hilbert space in the limit of $g \ll \eta$ is divided into bands, we truncate the Hilbert space to obtain an effective Hilbert space with a fixed number of bands. We denote the resulting truncations in the notation $[d,t]$, where $d$ is the maximun number of doublons (double excited states $\ket 2$) allowed, and $t$ is the  maximun number of ``triplons'' (triple excited states $\ket 3$) allowed. More explicitly, the effective Hilbert space $[1,0]$ includes the first two bands: the qubit subspace band, and the band with exactly one doublon among all the gmons. The total dimension of this effective Hilbert space is
\begin{align}
{N \choose  {N/2}} + {N \choose N/2-1,N/2-2,1}\;.
\end{align}
Analogously, the effective Hilbert space $[2,0]$ includes the first three bands: the qubit subspace band, the band with exactly one doublon among all the gmons, and the band with exactly two doublons among all the gmons. We will also consider the effective Hilbert space $[2,1]$, which includes four bands: the bands in the $[2,0]$ Hilbert space, and the band with exactly one triplon among all the gmons.

\begin{figure}%[H]
	\centering\includegraphics[width=\columnwidth]{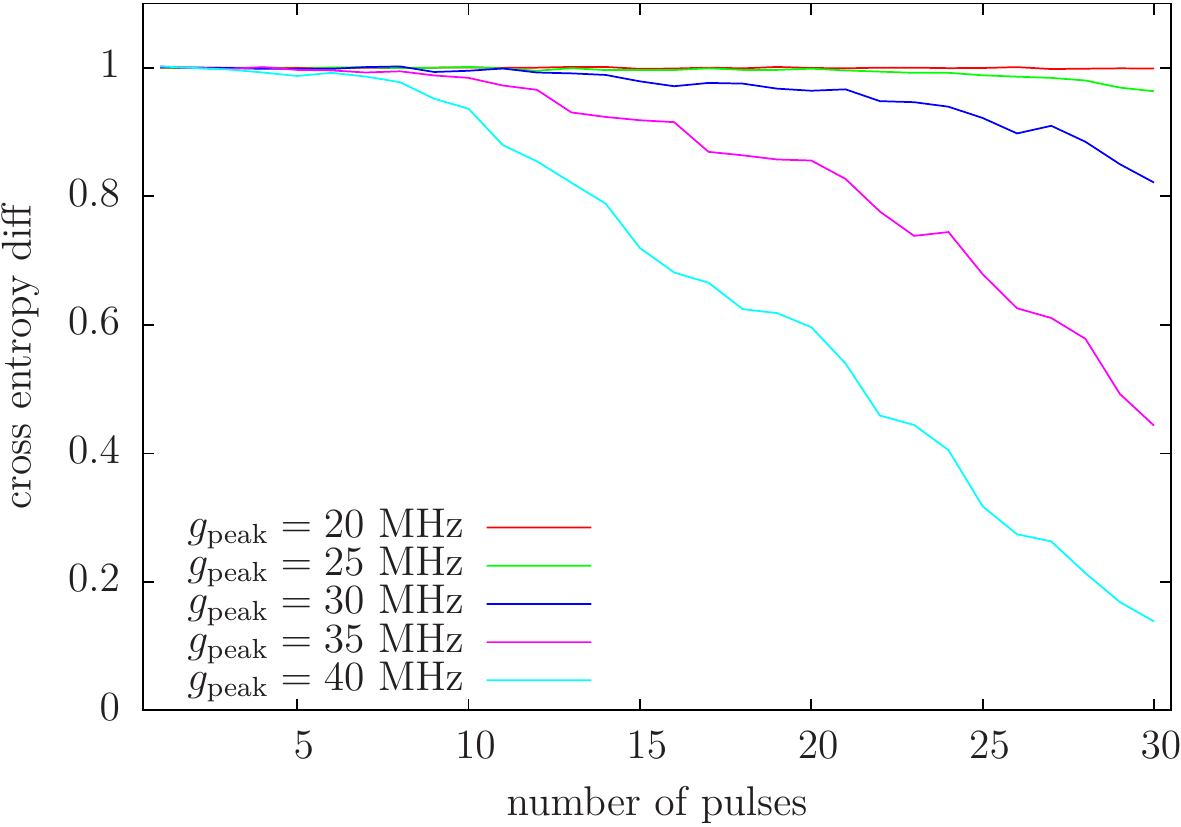}
	\caption{Cross entropy difference with four bands for 16 gmons, the effective Hilbert space $[2,1]$ (four bands) and $T_\text{pulse} = 20.5 \pm 4.5$ ns.}\label{fig:ce_4bands}
\end{figure}

\begin{figure}%[H]
	\centering\includegraphics[width=\columnwidth]{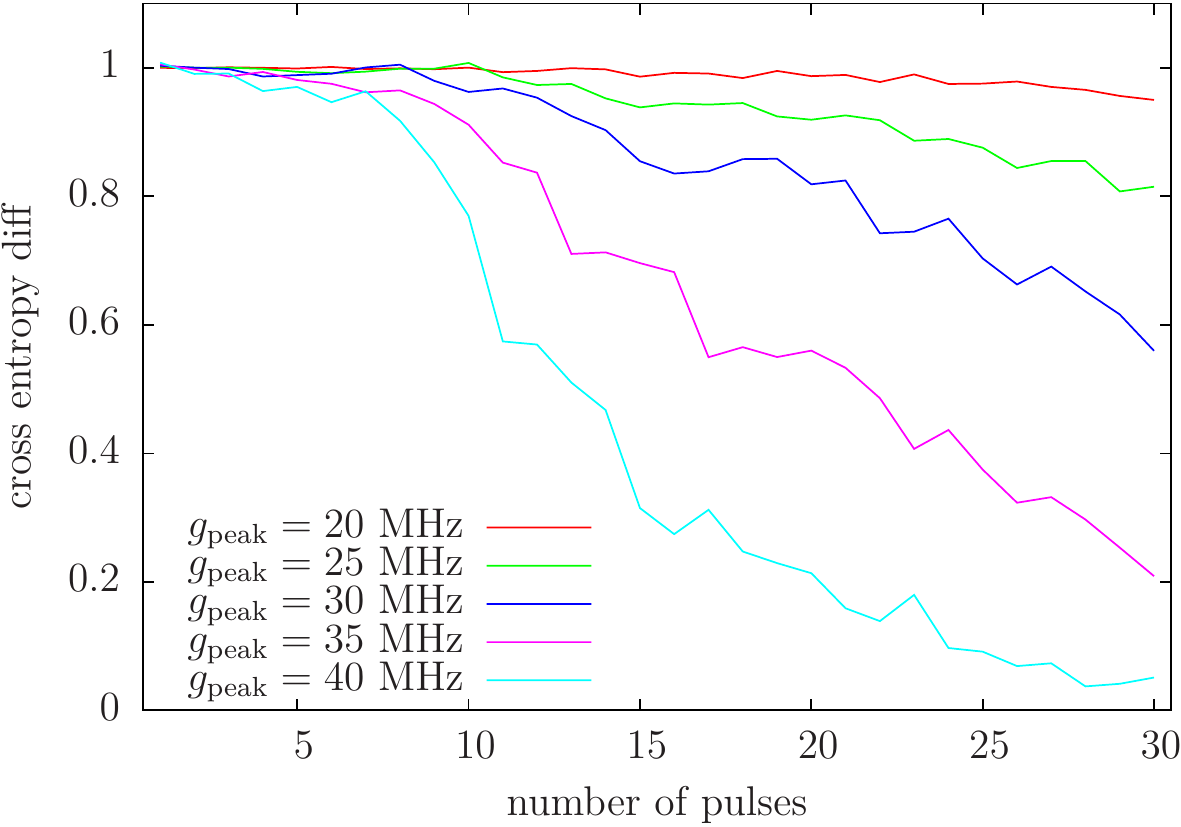}
	\caption{Cross entropy difference with three bands for 16 gmons, the effective Hilbert space $[2,0]$ (three bands) and $T_\text{pulse} = 20.5 \pm 4.5$ ns.}\label{fig:ce_3bands}
\end{figure}

\begin{figure}%[H]
	\centering\includegraphics[width=\columnwidth]{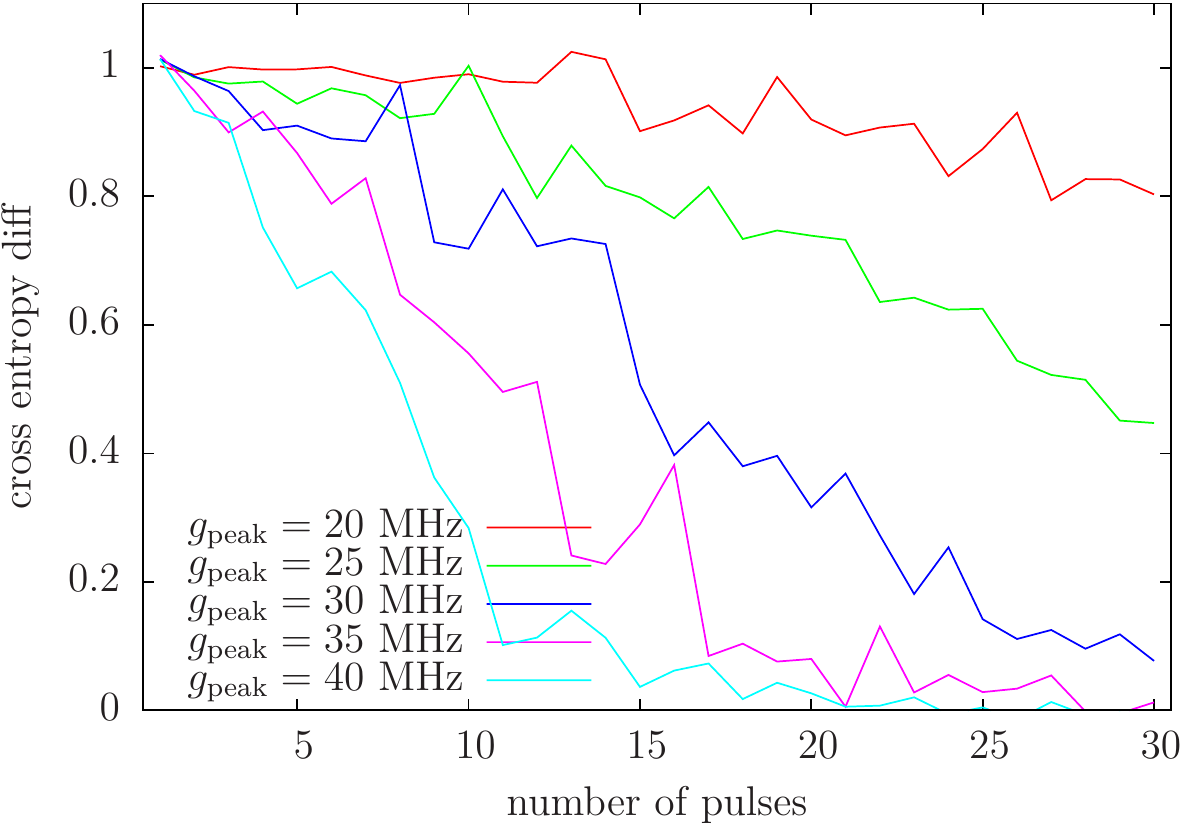}
	\caption{Cross entropy difference with two band for 16 gmons, the effective Hilbert space $[1,0]$ (two bands) and $T_\text{pulse} = 20.5 \pm 4.5$ ns.}\label{fig:ce_2bands}
\end{figure}

Figure~\ref{fig:ce_4bands} shows the cross entropy difference where we use the $[2,1]$ effective Hilbert space. As in Sec.~\ref{sec:app_direct_truncation}, we calculate the cross entropy difference comparing with the $m=4$ simulation which, as we have seen, is an accurate simulation. We observe that the numerical error resulting from the truncation is less than the expected experimental error, even for fairly large maximum $g$ and a substantial number of pulses. Therefore, this is a valid numerical method for the purpose of comparing the classical cost of simulation to a future experiment. Figure~\ref{fig:ce_3bands} shows the cross entropy difference where we use the $[2,0]$ effective Hilbert space. Although the errors are bigger than in the $[2,1]$ effective Hilbert space, this remains a valid numerical method. Finally, Fig.~\ref{fig:ce_2bands} shows the cross entropy difference where we use the $[1,0]$ effective Hilbert space. The numerical error is in this case larger, but probably still comparable to the expected experimental errors.

\subsubsection{Memory and time-estimate for simulations.}

\begin{figure}
	\centering
	\includegraphics[width = \columnwidth]{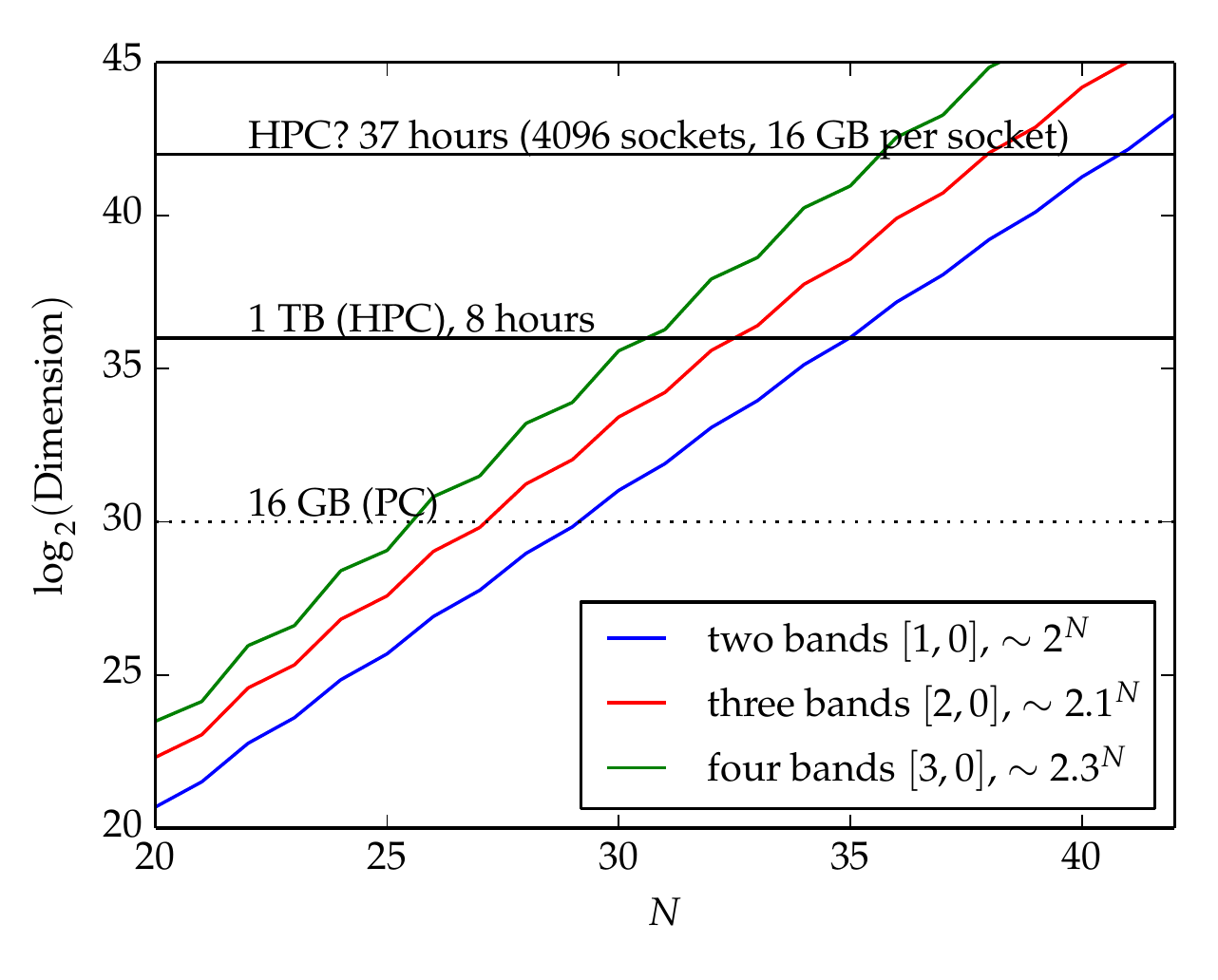}
	\caption{Estimates of memory and computational time for a numerical simulation using fourth order Runge-Kutta and an effective Hilbert space truncated to two bands (blue) and three bands (red).}\label{fig:mt_estimates} 
\end{figure}

We now provide estimates of the memory and computational time required for an approximate classical numerical simulation from the discussion given above. Figure~\ref{fig:mt_estimates} plots $\log_2({\rm D})$, where $D$ is the dimension of the effective Hilbert space, for increasing number of gmons $N$. We use this metric because, if we were studying a fully chaotic evolution in a Hilbert space composed of qubits, then  $\log_2({\rm D})$ would be the number of qubits. We plot the dimension of the effective space using two bands (the $[1,0]$ subspace in the notation of Sec.~\ref{sec:app_fixed_bands}), using three bands  (the $[2,0]$ subspace), and using four bands (the $[2,1]$ subspace). We see that for the number of gmons $N$ experimentally relevant in the near future, both curves are well approximated by exponentials. The effective dimension is then $D\sim 2^N$ for two bands (the $[1,0]$ subspace) $D\sim 2.1^N$ for three bands (the $[2,0]$ subspace), and $D \sim 2.3^N$ for four bands (the $[2,1]$ subspace). In the same figure we plot horizontal lines corresponding to 16 GB (and 1 TB) of memory required to store a quantum state with $\log_2(D)$ qubits, using a double precision complex number to store each amplitude, so that the memory for a quantum state is $D\times 16$ bytes. 

Estimating the corresponding run time is more subtle. For large scale simulations in classical supercomputers, the run time is mostly constrained by the network bandwidth between the different nodes~\cite{BoixoCircuits,haner20170}. If we store the quantum state in memory, distributed among multiple nodes in a supercomputer, a single integration step in fourth order Runge-Kutta would require approximately 5 swaps between nodes (10 vector communications, taking into account that the vector amplitudes need to go out and into each node). Therefore, we expect that the total computational cost would be bandwidth bound. We label in Fig.~\ref{fig:mt_estimates} two horizontal lines corresponding to prelimnary estimates of 8 hours and 37 hours of required communication time. To arrive at this estimates, we assume 5 memory swaps per Runge-Kutta step, and 1000 Runge-Kutta steps. For 1 TB of memory per state, we assume 64 sockets (nodes) and an effective 6 GB/s network bandwidth per socket. For $D = 2^{42}$, corresponding to 70 TB of memory per state, we assume 4096 sockets and an effective bandwidth of 1.2 GB/s per socket. These estimates are taken from the values reported in Refs.~\cite{BoixoCircuits,haner20170}.

\subsection{Signatures of chaos}

\subsubsection{Quantum loss of memory}\label{sec:chaotic_state}

\begin{figure}%[H]
	\centering\includegraphics[width=\columnwidth]{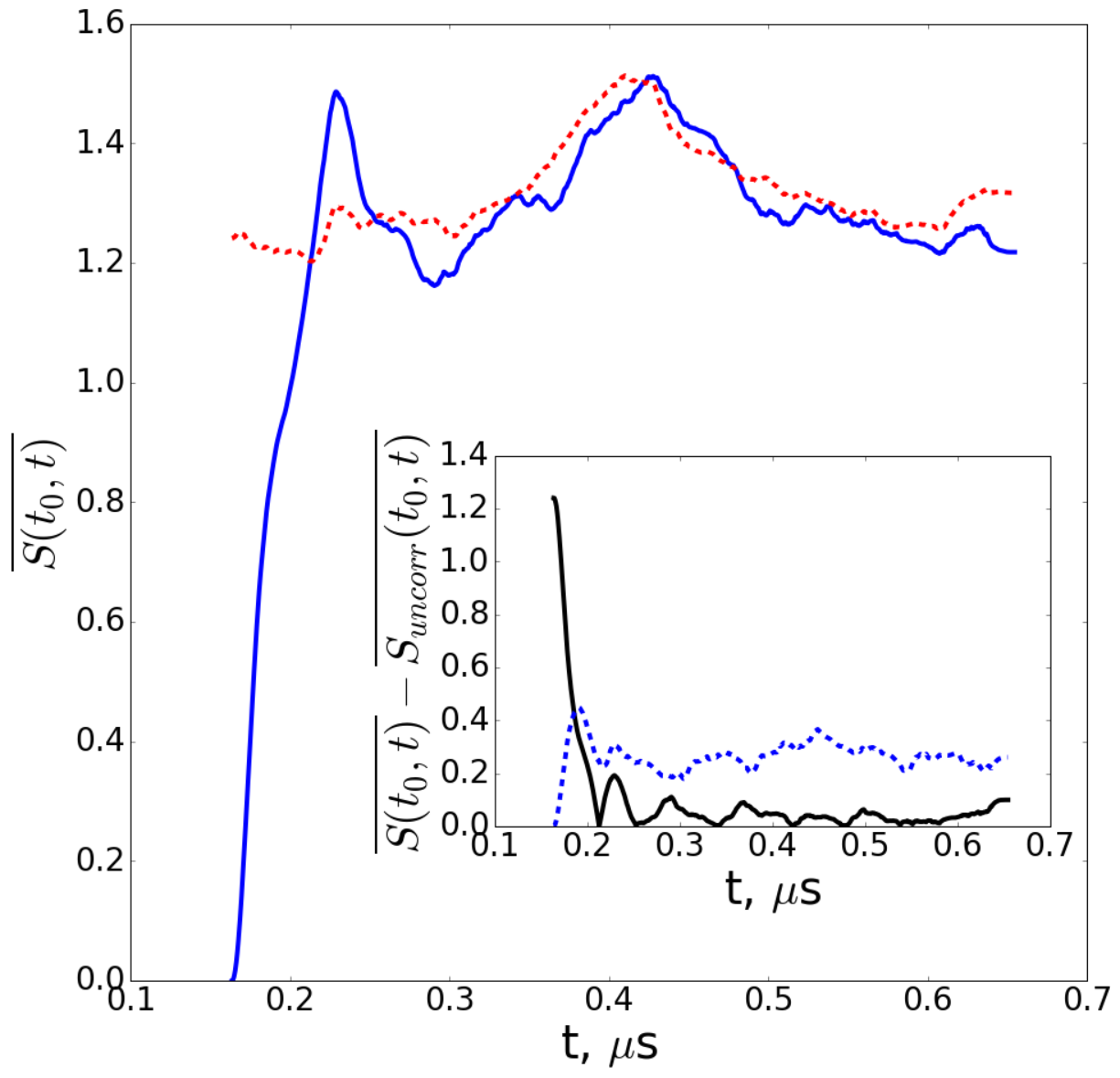}
	\caption{Time-cross-entropy averaged over random detuning $\delta_i$ and pulse shape. The parameters in the plot are $N=10$, $\delta =\pm 5\text{MHz}$,  $22.4\text{MHz}\leq \mathrm{max}\{g\} \leq 38.4\text{MHz}$, with variable pulse length $0.042\mu s \leq T_{\rm pulse} \leq 0.072\mu s$. Inset shows the deviation of the time-cross-entropy from the limit of uncorrelated time points, solid black line. The dashed blue line shows the  root mean square of the time-cross-entropy over the ensemble of disorder realizations.}\label{fig:TimeCrossEntropy}
\end{figure}

The wave functions and the spectrum of the uniform Bose-Hubbard model can be obtained via Bethe-Ansatz. This model has a regular spectrum characterized by conserved Bethe numbers~\cite{NagaosaBook}. In presence of disorder Bethe numbers are no longer integrals of motion and the system demonstrates quantum chaotic dynamics.  
%(ii) unitary evolution gives rise to statistics approximating Porter-Thomas distribution indicative of uniform sampling from Haar measure
As a result of chaotic dynamics the wave function amplitude  spreads uniformly over the available Hilbert space. A strong indication of chaos is the Porter-Thomas wave function statistics characteristic of Haar measure. 
%(iii) there is an analog of the Lyapunov exponent  for the system 
In practice the Porter-Thomas distribution is approached only approximately in the course of the shallow depth evolution (see Sec.~\ref{sec:pt_convergence}) and therefore it is important to check the consistency of the circuit dynamics with quantum chaos more carefully. Another characteristic of chaos is the rapid loss of memory of the initial state as a function of time. We characterize this phenomenon using statistics of the wave function over time. Consider the likelihood to observe bitstrings $z_1,\text{...},z_m$ at time $t$ in the course of the evolution,
\begin{align}
-\text{Log} \(\mathcal{L}_m(t)\)&\equiv -\text{Log}\left(\prod _{i=1}^m p_{z_i}(t)\right)\\&=-\sum _{i=1}^m\text{Log}\left(p_{z_i}(t)\right).
\end{align}
The likelihood has  Gaussian distribution centered at $\overline{\text{Log} \mathcal{L}_m}\approx  m \sum_z p_z  \log \left( p_z \right)$ with width of order $\sqrt{m}$. The loss of memory can be characterized by the ratio of likelihoods obtained using statistics at two different points in time,
\begin{align}
S\left(t_0,t\right)&\equiv-\lim_{m\rightarrow\infty}\frac{1}{m}
\text{Log} \left( \frac{\mathcal{L}_m(t)}{\mathcal{L}_m\left(t_0\right)}\right)
\\&\approx
- \sum
_zp_z\left(t_0\right)\text{Log}\left(\frac{p_z(t)}{p_z\left(t_0\right)}\right);.
\end{align}
This is the Kullback-Leibler divergence between two different times. The advantage of this characteristic is that it is directly measurable in our experiment and it quickly converges to its average for a sufficiently large number of measurements $m$. For perfectly correlated statistics as $t\rightarrow t_0$ the time-cross-entropy vanishes, $S(t_0,t_0)=0$. In a chaotic system we expect the statistics of the wave function at sufficiently  different moments in time to be fully uncorrelated,
\begin{multline}
\overline{S\left(t_0, t\gg t_0\right)} \approx \overline{S_{\text{uncorr}}}\left(t_0,  t\right) \\ \equiv -
\sum_z 
\left( 
\overline{p_z\left(t_0\right)} \;
\overline{ \text{Log}\left(p_z(t)\right)} 
-
\overline{p_z\left(t_0\right) \text{Log}\left(p_z(t_0)\right)} \nonumber
\right),
\end{multline}
Fig.~\ref{fig:TimeCrossEntropy} shows the cross-entropy $S(t_0,t)$ between $t_0$, taken after the first two pulses, and a later time $t$. $S(t_0,t)$ quickly deviates from zero (fully correlated) and approaches the uncorrelated value.

\subsubsection{Entanglement spread in the driven system}

\begin{figure}[h]
	% \centering
	\includegraphics[width=\columnwidth]{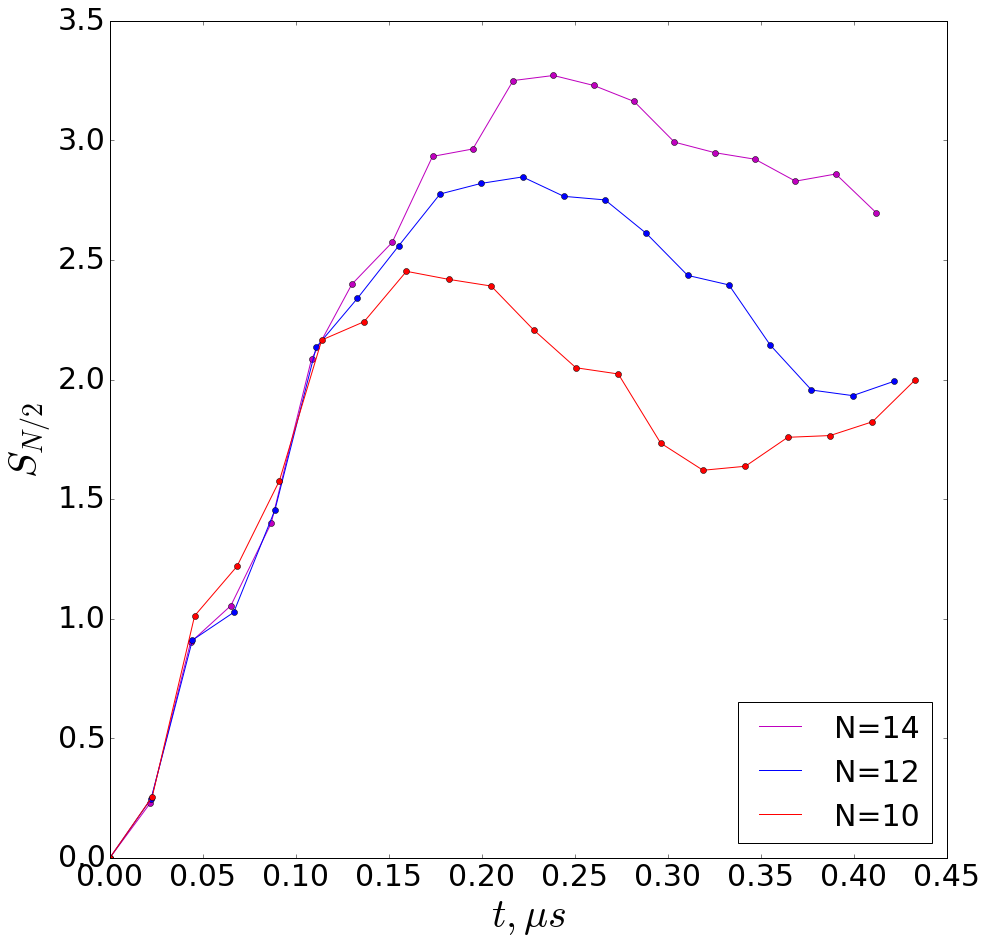}
	%\protect\protect
	\caption{Average entanglement entropy of the half-chain as a function of the system size. Parameters are $N=10$, $\delta =\pm 5\text{MHz}, \; 22.4\text{MHz}\leq \mathrm{max}\{g\} \leq 38.4\text{MHz}$, pulse length $0.042\mu s \leq T_{\rm pulse} \leq 0.072\mu s$.}
	\label{fig:EEvsN}
\end{figure}

\begin{figure}[h]
	% \centering
	\includegraphics[width=\columnwidth]{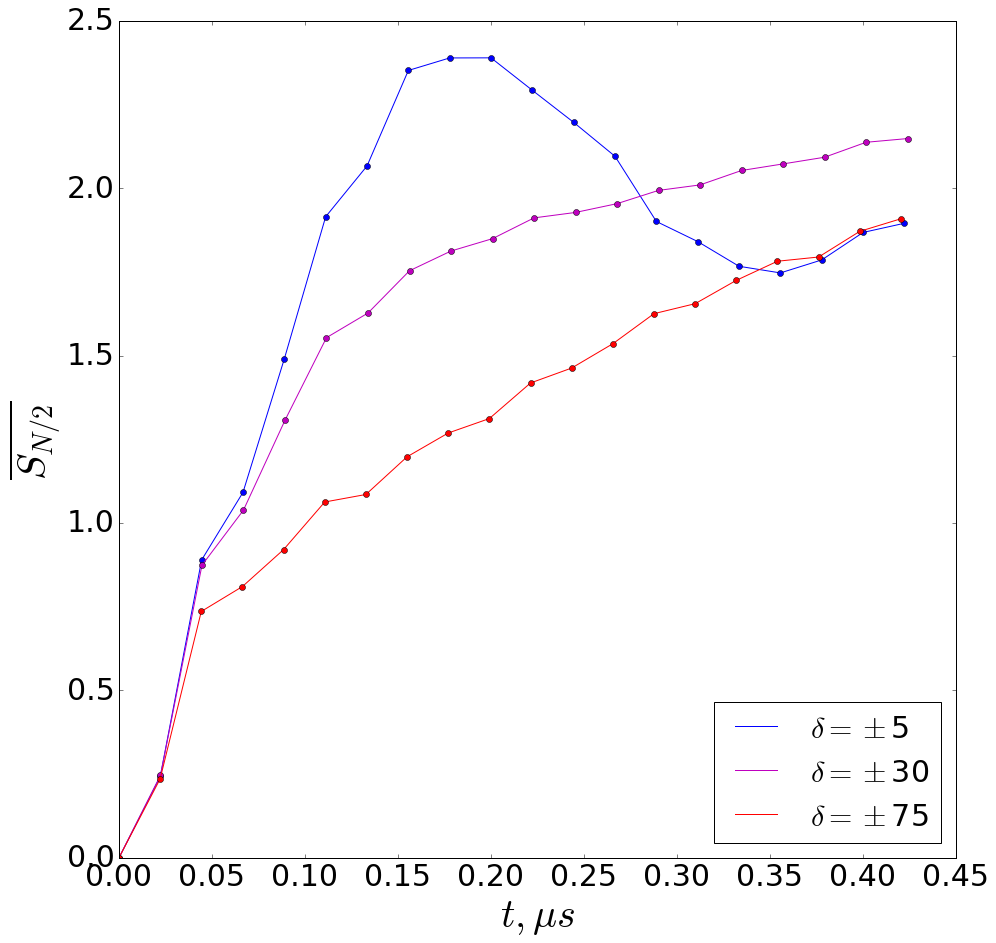}
	%\protect\protect
	\caption{Disorder strength. Same parameters as Fig.~\ref{fig:EEvsN}.}
	\label{fig:EEvsDisorder}  
\end{figure}

Quantum entanglement is an important, albeit not definitive~\cite{Knill2008}, metric of classical simulability of a quantum system. This is especially true in 1D as MPS and DMRG type algorithms have a computational cost exponential in the entanglement entropy. We analyze the bipartite entanglement entropy of a half chain,
\begin{gather}
S_{N/2 } \equiv \text{Tr}\left\{\hat{\rho }_{N/2} \text{Log}\left(\hat{\rho }_{N/2}\right)\right\},
\end{gather}
where $\hat{\rho }_{N/2}$ is the reduced density matrix. A random state from the Haar measure is expected to demonstrate volume law entanglement. In our experimental protocol we need to make sure the evolution time is long enough for the entanglement to achieve the regime of volume law scaling. The character of the entanglement spread depends on the relative strength of disorder and the spectral characteristics of the drive. At low disorder entanglement spreads ballistically as expected in the ergodic phase~\cite{AbaninEntanglement2017}, even in presence of diffusive particle transport~\cite{KimHuse2013}. Fig.~\ref{fig:EEvsN} shows the entanglement entropy $S_{N/2 }$ as a function of time for different system sizes, for a pulse sequence similar to the one used in the experiment. The velocity of entanglement spread appears to be independent of the system size suggesting that for the specific drive protocol and the initial state chosen the dynamics is similar to particle transport in a spin model.

At stronger disorder there is a crossover to the regime of sublinear spread of entanglement. In general, a periodically driven system undergoes a transition to the Many-Body localized phase~\cite{bordia_periodically_2017}, a regime in which it does not absorb energy. In contrast, in our protocol the drive consists of a wide range of harmonics and therefore we do not expect a many-body localized phase and the system continues to absorb energy. The growth of entanglement entropy however slows down dramatically at strong disorder, see Fig.~\ref{fig:EEvsDisorder}, likely due to the effect of rare regions with strong fluctuations of disorder potential which have a significant effect on entanglement transport in 1D~\cite{HuseGriffithsEffects,HuseRareRegions1D2017}.

\subsubsection{Convergence to Porter-Thomas}\label{sec:pt_convergence}

\begin{figure}%[H]
	\centering\includegraphics[width=\columnwidth]{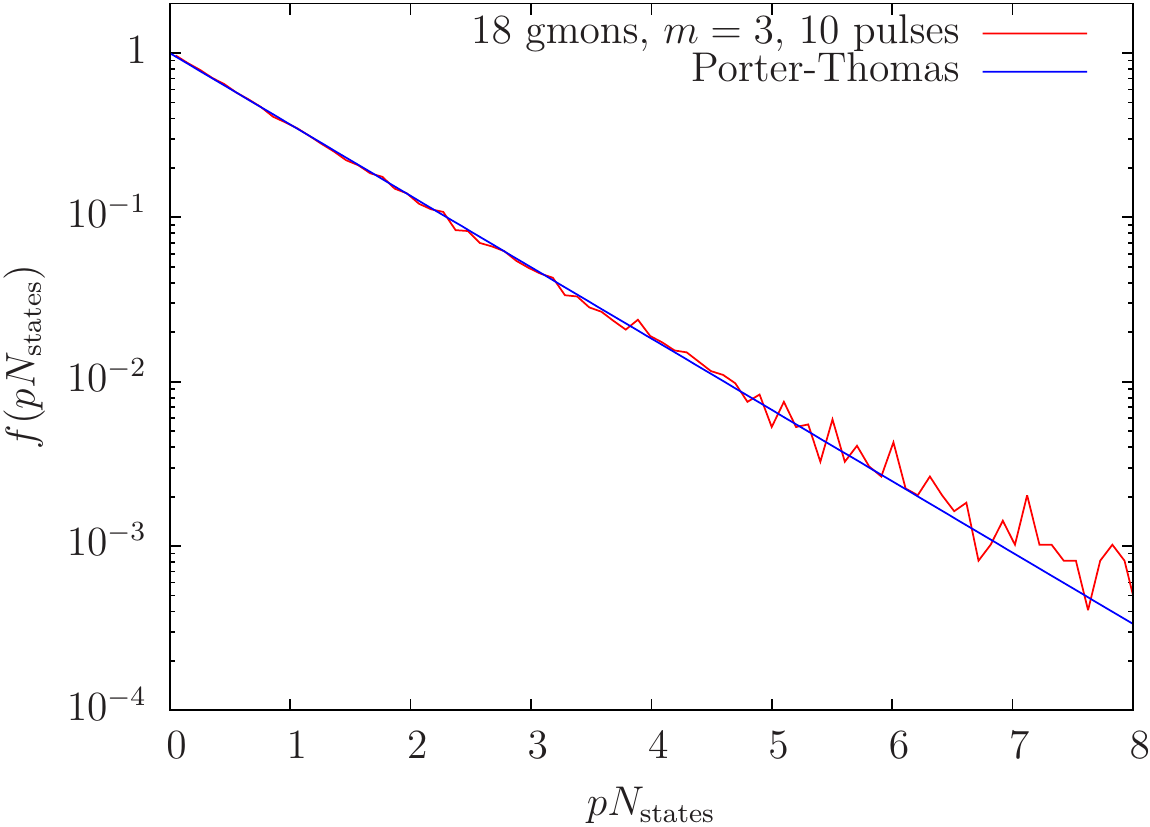}
	\caption{Histogram of output probabilities for 18 gmons and 10 pulses of duration $T_{\rm pulse} = 30 \pm 10$.}\label{fig:hist_m3}
\end{figure}

\begin{figure}%[H]
	\centering\includegraphics[width=\columnwidth]{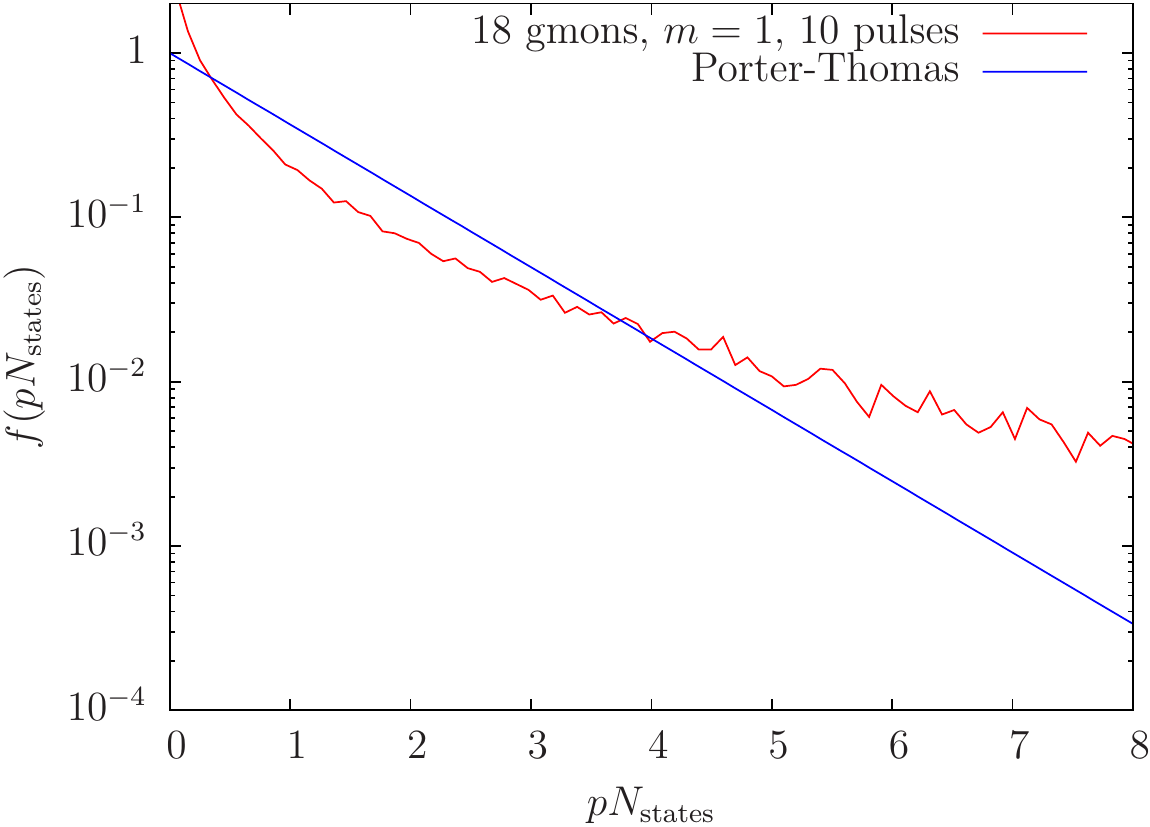}
	\caption{Histogram of output probabilities for 18 gmons, using plain qubits, and 10 pulses.}\label{fig:hist_m1}
\end{figure}

We now study the convergence to the Porter-Thoomas (exponential) distribution. Figure~\ref{fig:hist_m3} shows the histogram of the output probabilities $p = \{p_z\}$ for 18 gmons after 10 pulses, at half filling. We project the output state in the qubit subspace. The Porter-Thomas distribution is the exponential function, $f(p N_{\rm states}) = e^{-p N_{\rm states}}$, where $N_{\rm states}$ is the dimension of the qubit subspace at half filling, $N_{\rm states}={N \choose N/2}$. We see a good approximation of the numerical output distribution to Porter-Thomas. Figure~\ref{fig:hist_m1} shows the histogram of the output probabilities $p = \{p_z\}$ for 18 gmons after 10 pulses, but with the evolution carried out using the approximation to free fermions, as explained in  Sec.~\ref{sec:free}.  We see that in this case the distribution does not approximate the Porter-Thomas distribution, and is not chaotic (see Sec.~\ref{sec:chaotic_state}).

\begin{figure}%[H]
	\centering\includegraphics[width=\columnwidth]{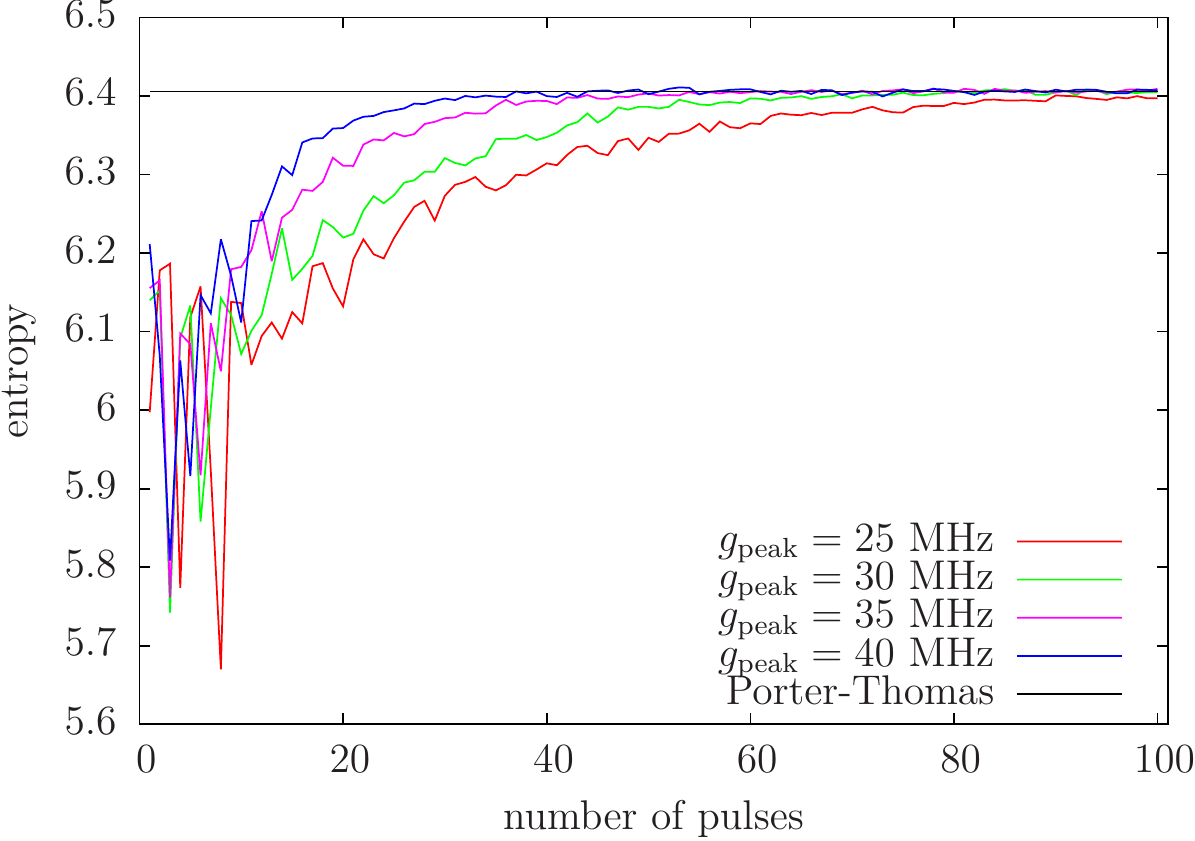}
	\caption{The entropy of the output distribution approaches Porter-Thomas with increasing number of pulses for 12 gmons, $m=3$ and $T_\text{pulse} = 20.5 \pm 4.5$ ns.}\label{fig:entropy_m3}
\end{figure}

\begin{figure}%[H]
	\centering\includegraphics[width=\columnwidth]{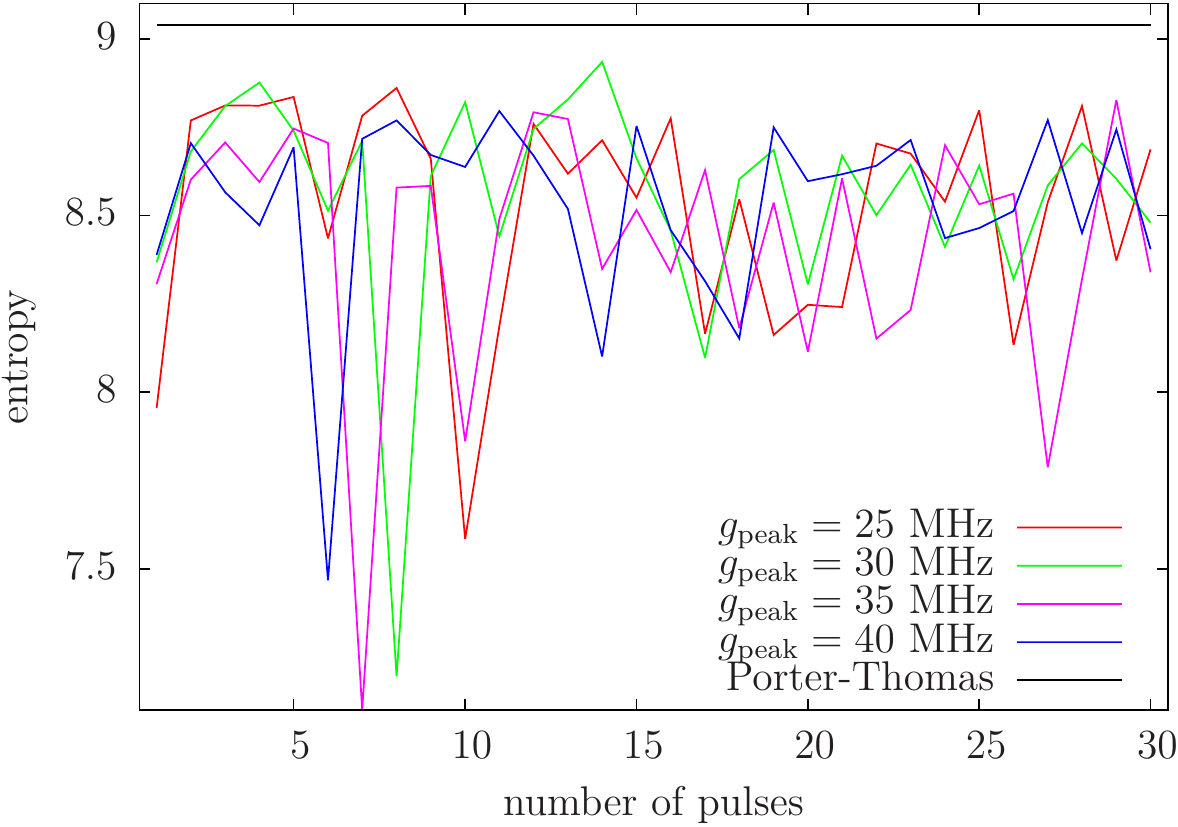}
	\caption{The entropy of the output distribution with plain qubits does not approach Porter-Thomas with increasing number of pulses for 16 gmons, $m=1$ and $T_\text{pulse} = 20.5 \pm 4.5$ ns.}\label{fig:entropy_m1}
\end{figure}

We depict in Fig.~\ref{fig:entropy_m3} the entropy of the output distribution with $N=12$ gmons as a function of the number of pulses and different maximun values of $g$ for the $g$-pulses. We see that it approaches the Porter-Thomas distribution entropy (black line) as the number of pulses increases. Figure~\ref{fig:entropy_m1} shows the entropy of the output distribution with increasing number of pulses using the approximation to free fermions with $N=16$ gmons. We see again that in this case the entropy does not converge to the Porter-Thomas entropy. 

\newpage
\bibliographystyle{apsrev4-1}